\documentclass[preprint,review,12pt]{elsarticle}
\biboptions{sort&compress}

\usepackage{amssymb}
\usepackage{amsmath}

\usepackage{float}
\usepackage{upgreek}
\usepackage{bm}
\usepackage{multirow}
\usepackage{booktabs}
\usepackage[margin=1.2in]{geometry}

\journal{Journal of Computational Physics}

\begin{document}

\begin{frontmatter}



\title{A Physics-Informed Machine Learning Framework for Solid Boundary Treatment in Meshfree Particle Methods} 


\author[label1]{Nariman Mehranfar} 
\author[label1]{Ahmad Shakibaeinia} 

\affiliation[label1]{organization={Department of Civil, Geological, and Mining Engineering, 
  Polytechnique Montreal},
            addressline={}, 
            city={Montreal},
            country={Canada}}

\begin{abstract}
Meshfree particle methods such as Smoothed Particle Hydrodynamics (SPH) and the Moving Particle Semi-Implicit (MPS) method are widely used to simulate complex free-surface and multiphase flows. A key challenge in these methods is the treatment of solid boundaries, where kernel truncation causes errors and instabilities. Traditional treatments, such as ghost particles and semi-analytical wall corrections, restore kernel completeness but add significant computational cost and complexity, especially for irregular geometries. We propose a physics-informed machine learning (ML) framework that directly predicts boundary correction terms for particle approximations, eliminating the need for ghost particles or analytical corrections. The framework is based on a hybrid convolutional neural network–multilayer perceptron (CNN–MLP) trained on physics-informed features that capture local geometry, particle states, and kernel properties. Once trained, it provides consistent boundary contributions across all spatial differential operators, including gradients, divergences, and Laplacians. The approach is demonstrated with MPS but is readily extensible to other particle methods such as SPH. Tests with predefined fields, unsteady diffusion, and incompressible Navier–Stokes flows show an accuracy comparable with ghost-particle methods while reducing computational overhead. The model generalizes well to unseen geometries, flow conditions, and particle distributions, including dynamically evolving domains. This work establishes a flexible, physics-informed ML paradigm for boundary treatment in particle-based PDE solvers, improving both accuracy and scalability across a broad class of meshfree methods.

\end{abstract}










\end{frontmatter}




\section{Introduction}
Meshfree Lagrangian (particle) methods such as Smoothed Particle Hydrodynamics (SPH) \cite{Gingold1977, Lucy1977}, and Moving Particle Semi-implicit (MPS) \cite{Koshizuka1996} are well-suited for fluid simulations, especially in problems involving large deformations and interface fragmentation. These methods represent the domain with freely moving particles, where the governing equations are approximated through a kernel smoothing process. SPH and MPS share many similarities, but they differ primarily in how spatial derivatives are approximated. In SPH, derivatives are computed by differentiating the kernel function, whereas in MPS, the kernel itself is used as a weight function, and the derivative operator is applied directly to the field variable.

A major challenge in both SPH and MPS is the treatment of solid boundaries, which strongly affects their accuracy, stability, and overall reliability. Near solid boundaries, the kernel support domain is truncated, leading to incomplete summation of particle contributions. This causes density deficiencies and spurious pressure gradients toward the boundary, which can lead to particle penetration to the boundary. Kernel truncation also increases approximation errors, especially for derivative estimates. In SPH, neglected surface-integral terms in the approximation of derivatives further contribute to errors near boundaries. While in MPS, derivatives approximation using directionally weighted differences mitigates this error, it still suffers from incomplete or asymmetric neighbor contributions. These issues directly impact the consistency and conservation properties of both methods. To address boundary-related errors, many approaches have been developed in the past, which broadly fall into three categories \cite{CercosPita2024, Ferrand2013, Zhang2019, Crespo2011, Harada2007, Price2011, Shadloo2016, Zhang2022, Xie2022}:

\begin{enumerate}
    \item Repulsive force methods that introduce short-range repulsive forces (e.g., Lennard–Jones-type) to prevent fluid particles from crossing the wall \cite{Monaghan2009, Crespo2007, Violeau2016}. They are simple to implement, but often suffer from poor physical fidelity and weak conservation properties \cite{Jinxin2023}.

    \item Ghost and dummy particle methods, which are among the most widely used approaches, populate the region outside the boundary with fictitious particles to restore kernel support. Particles can be placed in fixed predefined positions (commonly referred to as dummy particles) \cite{Koshizuka1996, Monaghan1994, Morris1997, Shakibaeinia2010, Vela2019} , or dynamically mirrored from nearby fluid particles (often called ghost particles) \cite{Harada2008, Adami2012, Lee2011, Duan2020, Colagrossi2003, Park2011, Akimoto2013}. As the literature uses these terms interchangeably, in this paper, we refer to all such particles as ghost particles for simplicity. While effective and physically consistent, these methods can be computationally expensive \cite{English2021}. The number of ghost particles can exceed that of fluid particles in large domains, significantly increasing memory and run-time costs. Defining and updating them is also challenging for complex geometries.

    \item Semi-analytical and hybrid approaches that explicitly correct for kernel truncation without relying on extra particles. Examples include semi-analytical wall boundary conditions (SAW) \cite{Mayrhofer2013, Kostorz2020}, and methods like the Unified Semi-Analytical Wall (USAW) \cite{Ferrand2013} and Polygon Wall (PW) boundary conditions \cite{Mitsume2015, He2019, He2018, Amaro2019, Zhang2017, Zhang2019, Zhang2014, Zhang2016, Mitsume2015}. While more direct and often accurate, they can also be computationally demanding, especially when complex analytical corrections are required \cite{Mitsume2015}. Approximations for complex geometries are also not always straightforward.

\end{enumerate}

Artificial intelligence (AI) and machine learning (ML) are increasingly being applied in computational science and engineering, including fluid mechanics, to complement traditional physics-based numerical models \cite{Brunton2020, Duraisamy2019, Raissi2019, Karniadakis2021, Brenner2019, Mahesh2019, Samuel1959, Alzubi2018, Chen2018}. Recent studies show that machine learning (ML) is being used in numerical modeling of PDEs in various ways. At one end of the spectrum, fully data-driven surrogate models can replace an entire solver \cite{Brunton2020, Willard2020}. In other cases, ML works alongside conventional numerical methods, with the goal of acceleration, enhancement, and optimization. For instance, ML can help approximate certain components of a PDE, like spatial derivatives or source terms, while the the numerical solver handles the rest, like in neural ODEs and discrete-time PINNs \cite{Raissi2019,Karniadakis2021}. Another approach is to use ML selectively for specific tasks (often computationally expensive or physically uncertain ones), such as learning turbulence closures, constitutive laws, or solving auxiliary algebraic equations like the pressure Poisson equation \cite{Duraisamy2019,Brunton2020}. ML is also commonly applied for parameter estimation, calibration, and model closure in physics-based simulations \cite{Duraisamy2019,Karniadakis2021}. Success of these examples illustrates how ML can help traditional physics-based modeling.

In recent years, particle methods have also increasingly benefited from machine learning (ML) techniques. Early work by \citet{Ladický2015} showed that particle positions and velocities could be learned using regression forests, while \citet{Marinho2021} employed k-nearest neighbors (kNN) to design an anisotropic SPH kernel. \citet{Bai2021} proposed a chained hashing algorithm for data-driven constitutive modeling in SPH. Neural networks have also been applied in different contexts—for example, \citet{Xiaoxing2021} used them to compute interface curvature in surface-tension models. More recently, physics-informed approaches have emerged, such as the Lagrangian formulations of physics-informed neural networks (PINNs) developed by \citet{Wessels2020} and \citet{Bai2021}.

Alongside these efforts, researchers have explored how particle neighbor lists can be leveraged as physics-informed aggregations of local information. \citet{Woodward2021} and \citet{Alexiadis2023} drew parallels between neighbor lists in particle-based systems and convolutional layers in grid-based data, highlighting their role in reducing computational complexity. \citet{Woodward2021} also developed reduced Lagrangian models for turbulence with varying levels of ML involvement, while \citet{Tian2022} extended ML applications to turbulence characterization. More recently, \citet{Zhang2023} replaced the pressure Poisson equation (PPE) in incompressible SPH (ISPH) with a CNN-based surrogate, achieving faster simulations without loss of accuracy. Despite this growing body of work, to the best of our knowledge, no ML framework has yet been developed specifically for boundary treatments in particle methods.

Building on recent progress in applying ML to numerical methods, this paper introduces a physics-informed, data-driven framework for solid boundary treatment in particle methods, demonstrated here for the MPS scheme. The goal is to develop a flexible and efficient alternative that avoids the complexity and cost of conventional approaches, such as ghost particles. The proposed model learns from ghost-particle data to predict boundary correction terms for each MPS operator, thereby replacing contributions previously computed by traditional treatments. To the best of our knowledge, this is the first ML framework specifically designed for boundary corrections in particle discretizations. The architecture employs a hybrid convolutional neural network–multilayer perceptron (CNN–MLP), which processes physics-inspired features—including geometric descriptors, field variables, and kernel/boundary properties, to predict boundary contributions. Separate models are trained for number density, gradient, divergence, and Laplacian operators. The training datasets cover diverse geometries and field conditions to promote generalization. The models can be considered physics-informed, as they learn from features and datasets derived from established physics-based methods, and they predict boundary contributions that remain consistent with the underlying physics of MPS.

The framework’s performance and generalization are assessed on three test cases with unseen geometries: (1) a spatially varying prescribed field with static particles; (2) a spatio-temporally varying field obtained from the solution of a pure diffusion PDE with static particles; and (3) a spatio-temporally varying field with dynamic particles governed by the Navier–Stokes and transport equations. In all cases, the ghost-particle approach provides the ground truth data. This study demonstrates the potential of physics-informed ML to replace conventional boundary treatments in MPS, offering a foundation for future extensions to three-dimensional problems and other particle-based methods.

\section{Methodology}

\subsection{MPS Particle Method}
The Moving Particle Semi-Implicit (MPS) method is a widely used meshfree particle scheme for incompressible flows, originally introduced by \citet{Koshizuka1996}. Here, we briefly summarize its standard formulations to establish the baseline for the proposed ML-based boundary treatment.

\subsubsection{Particle approximations}\label{sec:fundamentals}

\begin{figure}[H]
  \centering
  \includegraphics[width=0.3\textwidth]{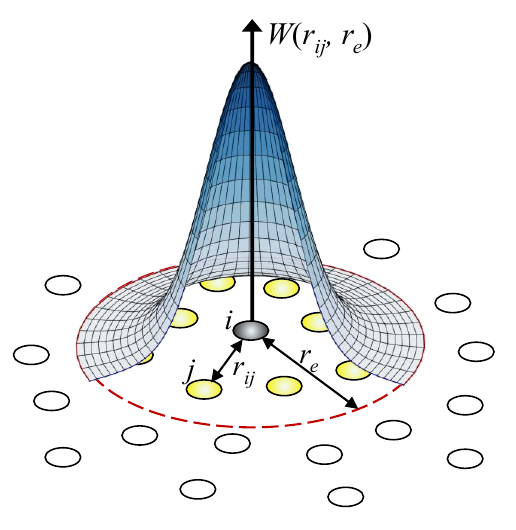}
  \caption[width=0.7\textwidth]{Kernel function in the 2-Dimensional MPS method}
  \label{Fig:Kernel}
\end{figure}

In MPS, similar to other particle methods, the continuum is represented by a set of freely moving nodes, referred to as particles, which carry physical quantities. The field variables and their derivatives are approximated through kernel-weighted interactions  between each particle target  $i$, with position vector $\mathbf{r}_i$, and its neighboring particles $j$, with position vector at $\mathbf{r}_j$. The interaction weighted by a kernel function $W_{ij} = W({r}_{ij}, r_e)$, where $r_{ij} =\|\mathbf{r}_{ij}\|= \|\mathbf{r}_j - \mathbf{r}_i\|$ is the relative position vector (distance), and $r_e$ is the kernel support radius (Fig. \ref{Fig:Kernel}) \cite{Koshizuka1996, Shakibaeinia2012}.

The kernel-weighted convolution of a scalar field $\phi$ is given by: :

\begin{equation}
\left\langle {\phi ({\bf{r}})} \right\rangle  = \frac{{\int_v {\phi \left( {{\bf{r'}}} \right){W_{ij}}dv'} }}{{\int_v {{W_{ij}}dv'} }}
\end{equation}
which in discrete form gives the MPS interpolation formula as \cite{Koshizuka1996}:
 
\begin{equation}
{\left\langle {\phi ({\bf{r}})} \right\rangle _i} = \frac{{\sum\limits_{j \ne i} {\left( {{\phi _j}{W_{ij}}} \right){V_j}} }}{{\int_v {{W_{ij}}dv} }} = \frac{{\sum\limits_{j \ne i} {\left( {{\phi _j}{W_{ij}}} \right)} }}{{\sum\limits_{j \ne i} { {{W_{ij}}}} }},{V_j} = \frac{{{m_j}}}{{{\rho _j}}}
\end{equation}

Here, $m_j$ is the mass of the particle $j$, and $\rho_j$ is its density. Unlike in the SPH method, in the MPS,  the approximations are normalized, so there is no requirement for a partition of unity, and the kernel function is dimensionless. The normalization factor, known as the particle number density, is defined as ${\left\langle n \right\rangle_i} = \sum\limits_{j \ne i} W_{ij}$, which serves as a dimensionless measure of local particle concentration. When the material density is constant (e.g., in incompressible flows), the particle number density can be replaced by a constant initial value $n_0$. Therefore, the interpolation operator can be written as:
\begin{equation}
{\left\langle {\phi ({\bf{r}})} \right\rangle _i} = \frac{1}{{{n_0}}}\sum\limits_{j \ne i} {\left( {{\phi _j}{W_{ij}}} \right)} 
\label{Eqn:4}
\end{equation}
The standard MPS approximation for spatial derivatives is based on kernel-weighted pairwise interactions between a target particle $i$ and each of its neighbors $j$. The MPS formulations for the gradient, divergence, and Laplacian operators (for scalar field $\phi$ and vector field $\bm{\upphi}$) are given by \citet{Koshizuka1996}:

\begin{equation}
\left\langle \nabla \phi \right\rangle_i = \frac{d}{n_0} \sum_{j \ne i} \left( \frac{\phi_{ij}}{r_{ij}} \mathbf{e}_{ij} W_{ij} \right)
\end{equation}

\begin{equation}
\left\langle \nabla \cdot \bm{\upphi} \right\rangle_i = \frac{d}{n_0} \sum_{j \ne i} \left( \frac{\bm{\upphi}_{ij}}{r_{ij}} \cdot \mathbf{e}_{ij} W_{ij} \right)
\end{equation}

\begin{equation}
\left\langle \Delta \phi \right\rangle_i = \left\langle \nabla^2 \phi \right\rangle_i = \frac{2d}{{n_0} \lambda} \sum_{j \ne i} \left( \phi_{ij} \right) W_{ij}
\end{equation}

\begin{equation}
\left\langle \Delta \bm{\upphi} \right\rangle_i = \left\langle \nabla^2 \bm{\upphi} \right\rangle_i = \frac{2d}{{n_0} \lambda} \sum_{j \ne i} \left( \bm{\upphi}_{ij} \right) W_{ij}
\end{equation}

where $\phi_{ij}=\phi_j - \phi_i$ in scalar field and $\bm{\upphi}_{ij}=\bm{\upphi}_j - \bm{\upphi}_i$ in vector field,  $d$  is the number of space dimensions, $\mathbf{e}_{ij}=\mathbf{r}_{ij}/r_{ij}$ is the unit direction vector between particles $i$ and $j$, and $\lambda$ is a normalization factor defined as:

\begin{equation}
\lambda = \left\langle r_{ij}^2 \right\rangle_i = \frac{\sum_{j \ne i} r_{ij}^3 W_{ij}}{\sum_{j \ne i} r_{ij} W_{ij}}
\end{equation}

Throughout this work, we adopt the third-order polynomial spiky function proposed by \citet{Shakibaeinia2010} as the kernel function:

\begin{equation}
W_{ij} = 
\begin{cases}
\left(1 - \frac{r_{ij}}{r_e} \right)^3, & \text{if } r_{ij} \le r_e \\
0, & \text{otherwise}
\end{cases}
\end{equation}

Note that the formulations presented above correspond to the standard MPS approximations. Over the years, various alternative formulations have been developed to improve conservation properties (e.g., \cite{Khayyer2011, Lee2011, Tamai2014, Jandaghian2020}). A comprehensive review of these developments can be found in \cite{Wu2025}. In this study, we primarily follow the original formulation to maintain consistency and facilitate comparison. The only exception is the adoption of the improved gradient formulation proposed by \citet{Jandaghian2020}, which is necessary for one of our test cases. It is given by:

\begin{equation}
\left\langle \tilde \nabla \phi \right\rangle_i = \frac{d}{n_0} \sum_{j \ne i} \left( \left( n_i \dfrac{\phi_j}{n_j} + n_j \dfrac{\phi_i}{n_i} \right) \mathbf{e}_{ij} W_{ij} \right)
\end{equation}

\subsubsection{MPS for flow and transport simulation}
Here, we briefly describe the application of the basic MPS formulations for simulating incompressible flow and heat transport, which is the phenomenon modeled in one of the test cases in this study. The governing equations consist of the mass and momentum conservation equations (Navier–Stokes), the advection-diffusion equation for heat transport. In the Lagrangian framework, advection terms in all equations are naturally absorbed, and a separate equation of motion is added. The governing equations are therefore given by \cite{Garoosi2020}:

\begin{equation}
\frac{1}{\rho} \frac{\mathrm{D} \rho}{\mathrm{D} t} + \nabla \cdot \mathbf{u} = 0
\label{eq:continuity}
\end{equation}

\begin{equation}
\frac{{\mathrm{D}\rho {\bf{u}}}}{{\mathrm{D}t}} =  - \nabla p + \mu {\nabla ^2}{\bf{u}} + \bf{f}_b
\end{equation}
\begin{equation}
\frac{{\mathrm{D}\rho {C_p}T}}{{\mathrm{D}t}} = k{\nabla ^2}T
  \label{Eqn:diffusion}
\end{equation}
\begin{equation}
\frac{{\mathrm{D}{\bf{r}}}}{{\mathrm{D}t}} = {\bf{u}}
\end{equation}

where $\mathbf{u}$ is the velocity vector, $t$ is time, $p$ is pressure, $\rho$ is density, $\mu$ is the kinematic viscosity, $T$ is temperature, $k$ is the thermal conductivity, $C_p$ is the specific heat capacity, and $\mathbf{f}_b$ represents the body force vector. Pressure is computed using a weakly compressible approach (WC-MPS)~\cite{Shakibaeinia2010}, which employs an explicit equation of state (EoS) to relate pressure to density, i.e., $p = f(\rho)$. 
The MPS discretization of the spatial derivatives in the governing equations is given by:
\begin{equation}
\left\langle \nabla p \right\rangle_i = \frac{d}{n_0} \sum_{j \ne i} \left( \frac{p_{ij}}{r_{ij}} \mathbf{e}_{ij} W_{ij} \right)
\label{eq:grad_p}
\end{equation}
\begin{equation}
\left\langle \nabla \cdot \mathbf{u} \right\rangle_i = \frac{d}{n_0} \sum_{j \ne i} \left( \frac{\mathbf{u}_{ij}}{r_{ij}} \cdot \mathbf{e}_{ij} W_{ij} \right)
\end{equation}
\begin{equation}
\left\langle \nabla^2 \mathbf{u} \right\rangle_i = \frac{2d}{n_0 \lambda} \sum_{j \ne i} \mathbf{u}_{ij} W_{ij}
\end{equation}
\begin{equation}
\left\langle \nabla^2 T \right\rangle_i = \frac{2d}{n_0 \lambda} \sum_{j \ne i} T_{ij} W_{ij}
\end{equation}

For the pressure gradient, one can use \citet{Jandaghian2020} alternative formulation:

\begin{equation}
\left\langle \tilde \nabla p \right\rangle_i =
\frac{d}{n_0} \sum_{\substack{j=1 \\ j \ne i}}^{N}
\left( n_i \frac{p_j}{n_j} + n_j \frac{p_i}{n_i} \right)
\frac{\mathbf{e}_{ij}}{r_{ij}} W_{ij}.
\label{eq:grad_p_extended}
\end{equation}

Time integration is performed using a predictor–corrector scheme. The predictor step calculates the predicted velocity,  $\mathbf{u}_i^*$, and  predicted position, $\mathbf{r}_i^*$, as:
\begin{equation}
\mathbf{u}_i^* = \mathbf{u}_i^k + \frac{\Delta t}{\rho_i} \left( \mathbf{f}_b + \mu \nabla^2 \mathbf{u}_i^k \right)
\end{equation}
\begin{equation}
\mathbf{r}_i^* = \mathbf{r}_i + \mathbf{u}_i^* \Delta t
\end{equation}

The corrected velocity,  $\mathbf{u}'$ ,  is calculated in the correction step as:
\begin{equation}
\mathbf{u}' = -\frac{\Delta t}{\rho_i} \nabla p^{k+1}
\end{equation}

Here, the superscript $k$ denotes the current time step. The updated velocity at the new time step is then given by:
\[
\mathbf{u}_i^{k+1} = \mathbf{u}_i^* + \mathbf{u}'
\]

The pressure is calculated using the equation of state as:
\begin{equation}
p_i^{k+1} = \frac{\rho c_0^2}{\gamma} \left[ \left( \frac{\left\langle n^* \right\rangle_i}{n_0} \right)^\gamma - 1 \right]
\end{equation}

In this expression, $\gamma$ is a constant (typically taken as 7), and $c_0$ is an artificial speed of sound. The predicted particle number density $\left\langle n^* \right\rangle_i$ is calculated based on the predicted positions $\mathbf{r}_i^*$.  

To improve the stability and accuracy of the method, we use an artificial density diffusion term in the continuity equation following the DMPS technique \cite{Jandaghian2020} and employ a particle regularization approach known as dynamic pairwise collision (DPC) \cite{Jandaghian2022}. 

\subsubsection{MPS Boundary treatment}
\begin{figure}[H]
  \centering
  \includegraphics[width=1\textwidth]{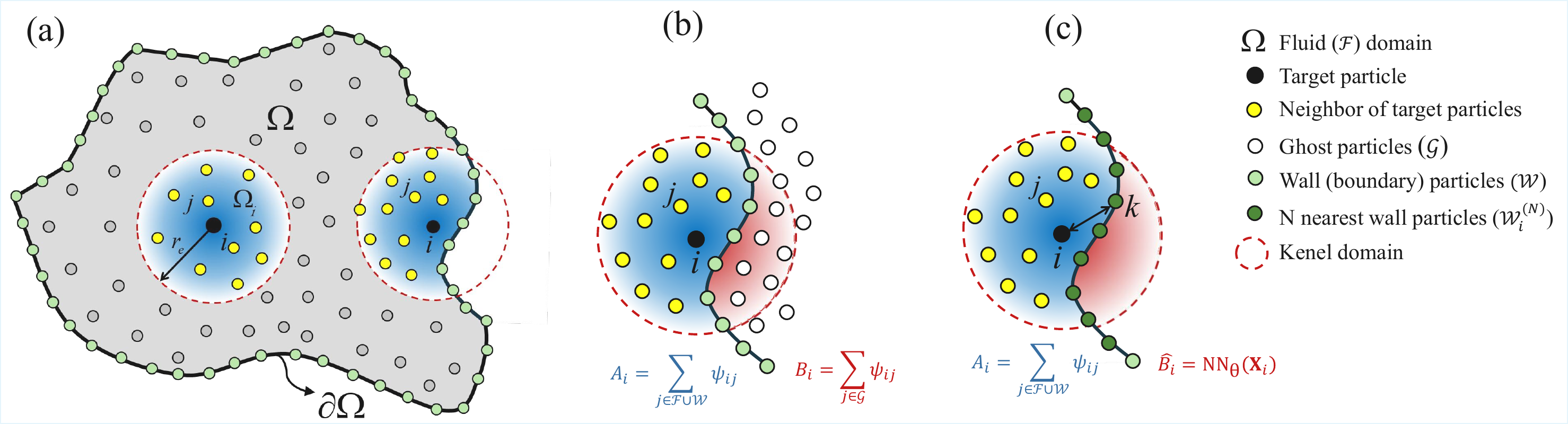}
  \caption[width=1\textwidth]{ Boundary treatment strategies in the MPS method. (a) Kernel truncation near boundaries. (b) Ghost particle method, which introduces artificial particles to compensate for the truncated kernel. (c) Proposed ML-based boundary treatment that predicts the missing boundary contribution.}
  \label{Fig:BC}

\end{figure}

As mentioned, boundaries pose challenges to the MPS method (similar to SPH) because the kernel support is truncated when a target particle is closer than $r_e$ to a boundary (Fig. \ref{Fig:BC}a). 
Several boundary treatment techniques have been developed to address kernel truncation and density deficiencies near boundaries in MPS (and similarly in SPH). The most widely used approach employs ghost/dummy particles, which populate the compact support across the boundary to restore kernel completeness and improve accuracy  (Fig. \ref{Fig:BC}b). Their physical properties (e.g., velocity, pressure) are either prescribed directly or extrapolated from neighboring interior particles, depending on the boundary condition type. Although ghost particle methods are effective, they struggle with treating complex boundaries and can have significant computational and memory overhead.

\subsection{Machine Learning for Boundary Conditions}
\subsubsection{Machine Learning Contribution in MPS}

Using machine learning instead of ghost particle methods in MPS boundary conditions offers several key benefits. It reduces the overall number of particles by eliminating the need for ghost particles, which leads to lower memory usage. Additionally, it simplifies the complexity of implementing boundary conditions by bypassing the need for complex ghost particle setups, making the method more flexible and easier to apply. 

To develop the method, let’s first express the MPS approximation of various operators ($\mathcal{L}$) applied to a scalar or vector field \( \phi \) over a target particle \( i \) in a general form as:
\begin{equation}
C_i = \langle \mathcal{L}\phi \rangle_i = \sum_{j \neq i} \psi_{ij},
\label{eq:ci_NN}
\end{equation}

where  \( \psi_{ij} \) defines the interaction between the particle $i$ and each of its neighbors $j$.  Table 1 provides the expression \( \psi_{ij} \) for the standard MPS approximation of various operators ($\mathcal{L}$), defined in section \ref{sec:fundamentals} , including for Number Density ($n$),  Interpolation ($\mathcal{I}$), Gradient ($\nabla$ and $\tilde\nabla$), Divergence ($\nabla \cdot$), and Laplacian ($\Delta$).

\begin{table}[H]
\centering
\caption{Expression $\psi_{ij}$ based on the standard MPS approximation of operators $\mathcal{L}$}
\resizebox{\textwidth}{!}{%
\begin{tabular}{|c|c|c|c|c|c|c|}
\hline
$\mathcal{L}$ & $n$ & $\mathcal{I}$ & $\nabla$ & $\tilde{\nabla}$ & $\nabla \cdot$ & ${\nabla^2}$ \\ \hline
$\psi_{ij}$ & 
$W_{ij}$ & 
$\dfrac{1}{n_0}\,\phi_j W_{ij}$ & 
$\dfrac{d}{n_0} \left( \dfrac{\phi_{ij}}{r_{ij}} e_{ij} \right) W_{ij}$ & 
$\dfrac{d}{n_0} \left( n_i \dfrac{\phi_j}{n_j} + n_j \dfrac{\phi_i}{n_i} \right) \dfrac{e_{ij}}{r_{ij}} W_{ij}$ & 
$\dfrac{d}{n_0} \left( \dfrac{\bm{\upphi}_{ij}}{r_{ij}} \cdot e_{ij} \right) W_{ij}$ & 
$\dfrac{2d}{n_0 \lambda} \phi_{ij} W_{ij}$ \\ \hline
\end{tabular}%
}
\end{table}

For the ghost particle approach, the formulation can be split into two components of \textbf{\( A_i \)} the contribution of the fluid (\( \mathcal{F} \)) and wall (\( \mathcal{W} \)) particles, and \textbf{\( B_i \)} the contribution of the ghost particles (\( \mathcal{G} \)), representing the boundary impact, expressed as:

\begin{equation}
\label{Eq:B}
C_i = \langle \mathcal{L}\phi \rangle_i = A_i + \underbrace{B_i}_{\text{boundary impact}} \quad \text{where} \quad A_i = \sum_{j \in \mathcal{F} \cup \mathcal{W}} \psi_{ij}, \quad \text{and} \quad B_i = \sum_{j \in \mathcal{G}} \psi_{ij}
\end{equation}

The goal of the machine learning  here will be to eliminate the ghost particles, and use a neural network $\mathrm{NN}_\theta \left( \mathbf{X}_i \right)$, with parameters $\theta$  and input features  $\mathbf{X}_i$ to predict a learnable correction term as boundary impact $\hat{B}_i$ (previously given by ghost particles contribution) as:

\begin{equation}
\label{Eq:B_hat}
\hat C_i = \left\langle \mathcal{L}\phi \right\rangle_i = A_i + \underbrace{\hat{B}_i}_{\text{boundary impact}} \quad \text{where} \quad A_i = \sum_{j \in \mathcal{F} \cup \mathcal{W}} \psi_{ij}, \quad \text{and} \quad \hat{B}_i = \mathrm{NN}_\theta \left( \mathbf{X}_i \right)
\end{equation}

Note that we keep the wall particles as they represent the boundary (see Fig. \ref{Fig:BC}c). The input feature vector \( \mathbf{X}_i \) is given by:

\begin{equation}
\mathbf{X}_i = \left\{
\left(
\mathbf{r}_i, \mathbf{r}_k, \mathbf{r}_{ik}, 
\lVert \mathbf{r}_{ik} \rVert, 
\lVert \mathbf{r}_{ik} \rVert^2, 
Nw_i, A_i, 
\phi_i, \phi_k, \phi_{ik}, 
\phi_{ik}  \mathbf{e}_{ik}, 
d_p, \frac{r_e}{d_p}, 
n_i, \lambda_i, 
\mathrm{BC}_k 
\right) 
\;\middle|\; k \in \mathcal{W}_i^{(N)}
\right\}
\end{equation}

and is defined to contain the following physics-inspired components:
\begin{enumerate}
    \item \textbf{Geometrical characteristics}, including:  
    the position of the target particle, $ \mathbf{r}_i $, the positions of the $ N $ nearest wall particles $ \mathbf{r}_k $ (for all \( k \in \mathcal{W}_i^{(N)} \), where \( \mathcal{W}_i^{(N)} \subset \mathcal{W} \)), the relative position of the target particle and its $ N $ nearest wall particles, $ \mathbf{r}_{ik} = \mathbf{r}_k - \mathbf{r}_i $, and its magnitude (i.e., distance) $ \lVert \mathbf{r}_{ik} \rVert $ and squared magnitude $\lVert \mathbf{r}_{ik} \rVert^2$, as well as the number of wall particles in the neighborhood (within support area) of $ i $, given by $ Nw_i = \sum_{\substack{ j \in \mathcal{W}}} \mathbb{I}(\lVert \mathbf{r}_{ij} \rVert^2 < r_e) $ where \( \mathbb{I}(\cdot) \) is the indicator function. Note that while $\mathbf{r}_{ik}$ already contains the information to derive $\lVert \mathbf{r}_{ik} \rVert$ and $\lVert \mathbf{r}_{ik} \rVert^2$, in practice this redundancy can helping learning by explicitly providing the model with useful nonlinearities included in the physics.

    \item \textbf{Field variable characteristics}, including:  
    contributions of fluid and wall particles, $ A_i $, field variable values at the target particle, $ \phi_i $, field variable values at the $ N $ nearest wall particles, $ \phi_k $, the difference $ \phi_{ik} = \phi_k - \phi_i $, and the direction product $ \phi_{ik}  \mathbf{e}_{ik} $.  Note that in the case of boundary contribution to particle number density ($n$), only $ A_i $  is required.

    \item \textbf{Kernel and boundary characteristics}, including:  particle size $ d_p $, normalized effective radius $ r_e / d_p $, particle number density $ n_i $, the coefficient in the Laplacian approximation formula $ \lambda_i $, and the boundary condition type $\mathrm{BC}_k$ (Dirichlet or Neumann). Through this study, we keep the kernel function type constant.
\end{enumerate}

\subsubsection{Machine Learning Architecture}
\begin{figure}[H]
    \centering
    \includegraphics[width=1\textwidth]{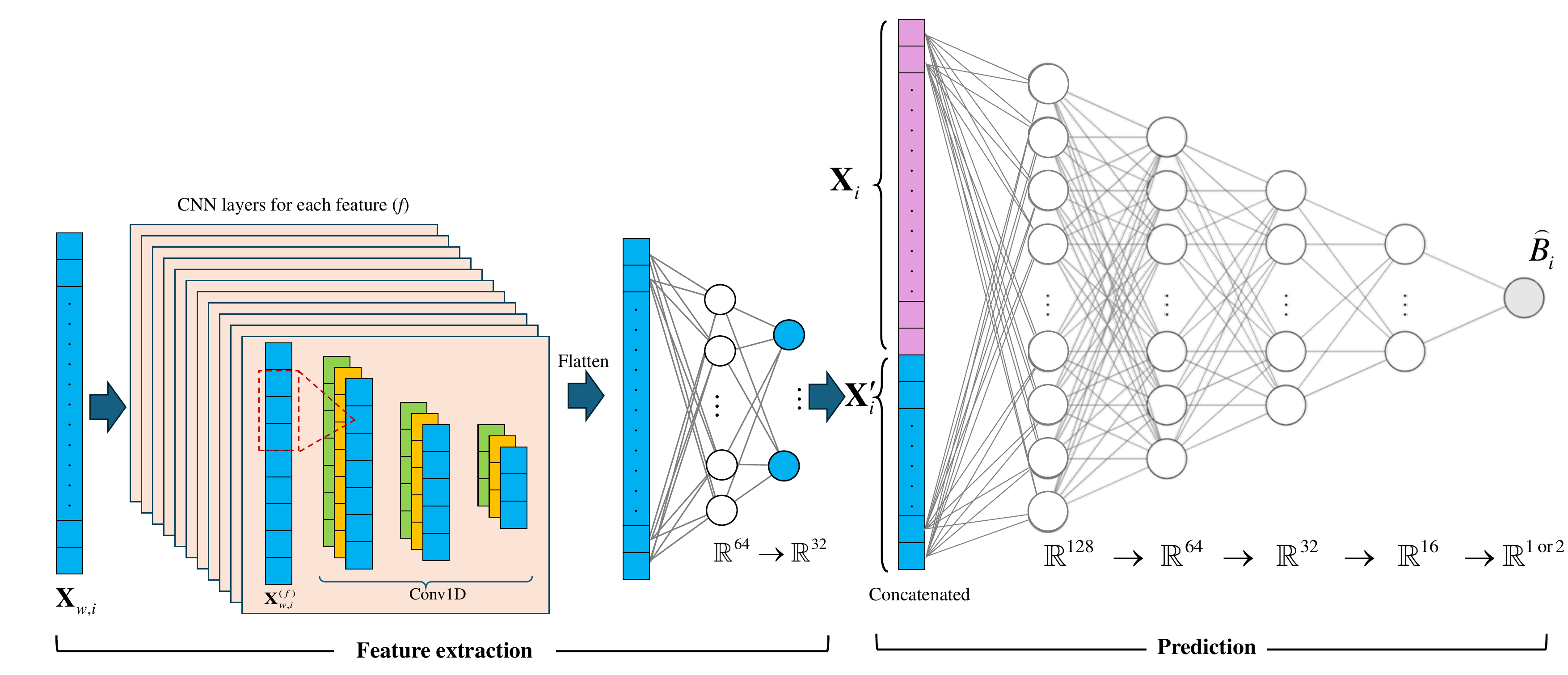}
    \caption[width=0.7\textwidth]{Schematic of Neural Networks Architecture}
    \label{Fig:Architecture}
\end{figure}

The machine learning model of this study includes a hybrid deep learning architecture that combines a feature extractor to capture local dependencies with a convolutional neural network (CNN) and a predictor (of boundary contribution $\hat{B}_i$) with a multilayer perception (MLP). Fig. \ref{Fig:Architecture} shows the architecture of this hybrid CNN-MLP network. Importantly, we found that MLP-only networks—even when made several times deeper than the hybrid design—failed to generalize beyond the training set, underscoring the need for the CNN-based feature extraction.

First, a subset of the input feature vector $\mathbf{X}i$ is extracted, corresponding to the $N$ nearest wall particles ($N = 9$ in this study, Fig. \ref{Fig:BC}c). This subset, denoted as $\mathbf{X}_{w,i}$, is used as input to the CNN to learn the geometric relationship between the wall particles and characterize the shape of the local boundary. 
CNN is particularly well-suited for this task due to its inherent ability to efficiently capture local patterns and spatial dependencies (defining the local boundary shape). Note that $N$ nearest wall particles are reordered so that they follow the natural progression along the boundary, required for CNN. The feature vector is defined as:

\begin{equation}
\mathbf{X}_{w,i} = 
\left\{ 
\left( \mathbf{r}_k, \mathbf{r}_{ik}, \lVert \mathbf{r}_{ik} \rVert, \lVert \mathbf{r}_{ik} \rVert^2, 
\phi_k, \phi_{ik}, \phi_{ik}  \mathbf{e}_{ik}, \mathrm{BC}_k \right) 
\;\middle|\; k \in \mathcal{W}_i^{(N)} 
\right\}
\end{equation}

For scalar fields $\phi$, this results in 11 features ($f$) per wall neighbor; for vector fields $\bm{\upphi}$, there are 12 features per neighbor. The CNN input is therefore a 1D vector 
$\mathbf{X}_{w,i} \in \mathbb{R}^{fN}$.  With $N=9$, this gives $fN=99$ for scalars and $fN=108$ for vectors. The CNN then processes this input by treating each feature, collected across the $N$ wall neighbors, as a separate 1D sequence $\mathbf{X}_{w,i}^{(f)} \in \mathbb{R}^N$. These sequences are passed through three 1D convolutional layers, each with three filters of size 3. The outputs are then flattened, resulting in a vector with the same length as $\mathbf{X}_{w,i}$.

This vector is further processed by two fully connected layers with 64 and 32 units, respectively. The resulting vector, denoted by $\mathbf{X}'_i$, is then concatenated with the full input feature vector $\mathbf{X}_i$ (a fusion of physics-inspired and data-driven inputs) and passed through a four-layer MLP with 128, 64, 32, and 16 units to predict the boundary contribution $\hat{B}_i$.
The length of $\mathbf{X}_i$  and $\mathbf{X}_{w,i}$ for each operator is summarized in Table~\ref{table:inputs}. As mentioned, for particle number density ${n}$, all features that containing field variable $\phi$ (or $\bm{\upphi}$ ) are excluded. The output layer of the network can be a vector or scalar, depending on the operation.

\begin{table}[H]
    \centering
    \caption{Lengths of input feature vectors used in the network}
    \label{table:inputs}
    \begin{tabular}{|c|c|c|}
        \hline
        Differential Operator $\mathcal{L}\phi$ & Length of $\mathbf{X}_i$ & Length of $\mathbf{X}_{w,i}$\\
        \hline
        $n$& 61 & 54\\
        $\left\langle \nabla \phi \right\rangle_i$ & 109 & 99\\
        $\left\langle \tilde\nabla \phi \right\rangle_i$ & 109 & 99\\
        $\left\langle \nabla \cdot \bm{\upphi} \right\rangle_i$ & 118 & 108\\
        $\left\langle \nabla^2 \phi \right\rangle_i$ & 108 & 99\\
        $\left\langle \nabla^2 \bm{\upphi} \right\rangle_i$ & 119 & 108\\
        \hline
    \end{tabular}
\end{table}

The activation function used in the feature extractor block (comprising the CNN layers and the first MLP) is \texttt{ELU}, while the predictor network (the second MLP) uses \texttt{ReLU}. All kernels are initialized using the \texttt{He normal} initializer. Biases are initialized with a \texttt{random normal} distribution having a mean of 0 and a standard deviation of $\sqrt{2 / \sigma}$, where $\sigma$ is the number of input units to the layer (see Table~\ref{table:inputs}). The output layer uses a \texttt{linear} activation function, with no special initialization for its kernel or bias.

The input features $\mathbf{X}_i$ and $\mathbf{X}_{w,i}$ are normalized using the mean and standard deviation computed from the training dataset. The network is trained using the \texttt{AdamW} optimizer with a scheduled learning rate and weight decay. The learning rate starts at $10^{-3}$ and decays to $10^{-5}$ over 1000 steps. The weight decay follows the same schedule, decreasing from $10^{-3}$ to $10^{-5}$ over the same period.

A custom loss function is employed to improve model convergence:
\begin{equation}
\text{Loss} = \text{MAE} + (1 - \max(R, 0))
\end{equation}

where MAE is the mean absolute error, and $R$ is the Pearson correlation coefficient given by:

\begin{equation}
\text{MAE} = \frac{1}{N} \sum_{i=1}^N \left| \hat B_i - B_i \right|
\end{equation}

\begin{equation}
R = \frac{ \sum\limits_{i=1}^N (B_i - \bar{B})(\hat{B}_i - \bar{\hat{B}}) }{ \sqrt{ \sum\limits_{i=1}^N (B_i - \bar{B})^2 } \sqrt{ \sum\limits_{i=1}^N (\hat{B}_i - \bar{\hat{B}})^2 } }
\end{equation}
Training is conducted over 2000 epochs, and the model achieving the best validation performance is selected as the final model, representing the cross-validation trade-off. This architecture was determined by testing various configurations of layer/unit counts for the two MLPs, and different activation functions including \texttt{ELU}, \texttt{ReLU}, \texttt{tanh}, \texttt{sigmoid}, \texttt{leaky ReLU}, and \texttt{swish}. The number and size of convolutional kernels were also optimized through experimentation.
Training is done using a batch size of 16,384 on a PC  with an \texttt{NVIDIA Tesla P40 (24 GB)} GPU, \texttt{AMD Ryzen 9 5950x} (16 cores, 32 threads) CPU, and 32~GB \texttt{DDR4-3200~MHz} RAM.

\subsection{Training dataset}
\begin{figure}[H]
    \centering
  \includegraphics[width=1\textwidth]{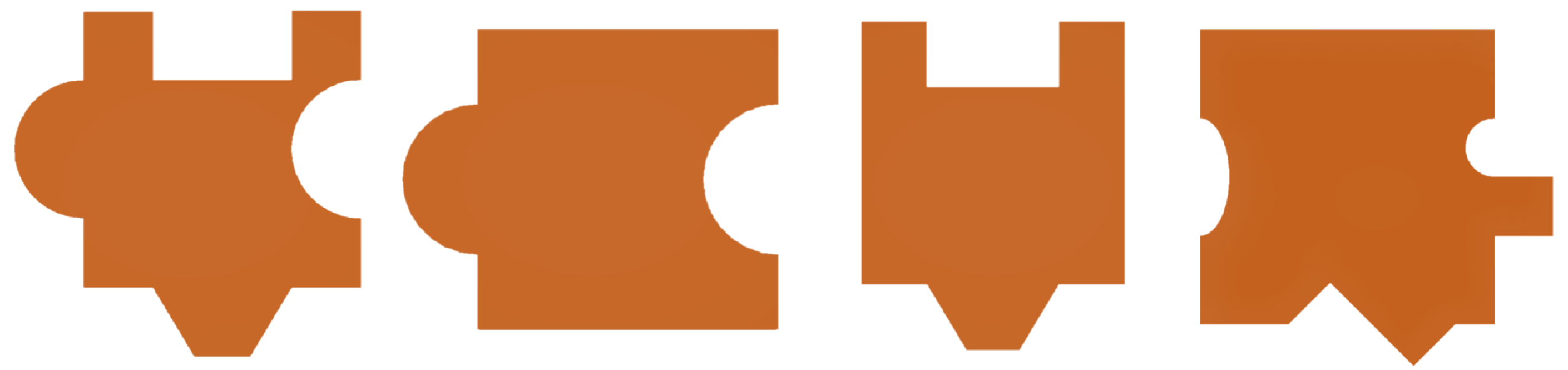}
    \caption[width=1\textwidth]{Geometries used for making the training set}
\label{Fig:Geometries}
\end{figure}

\begin{figure}[H]
  \centering
  \includegraphics[width=1\textwidth]{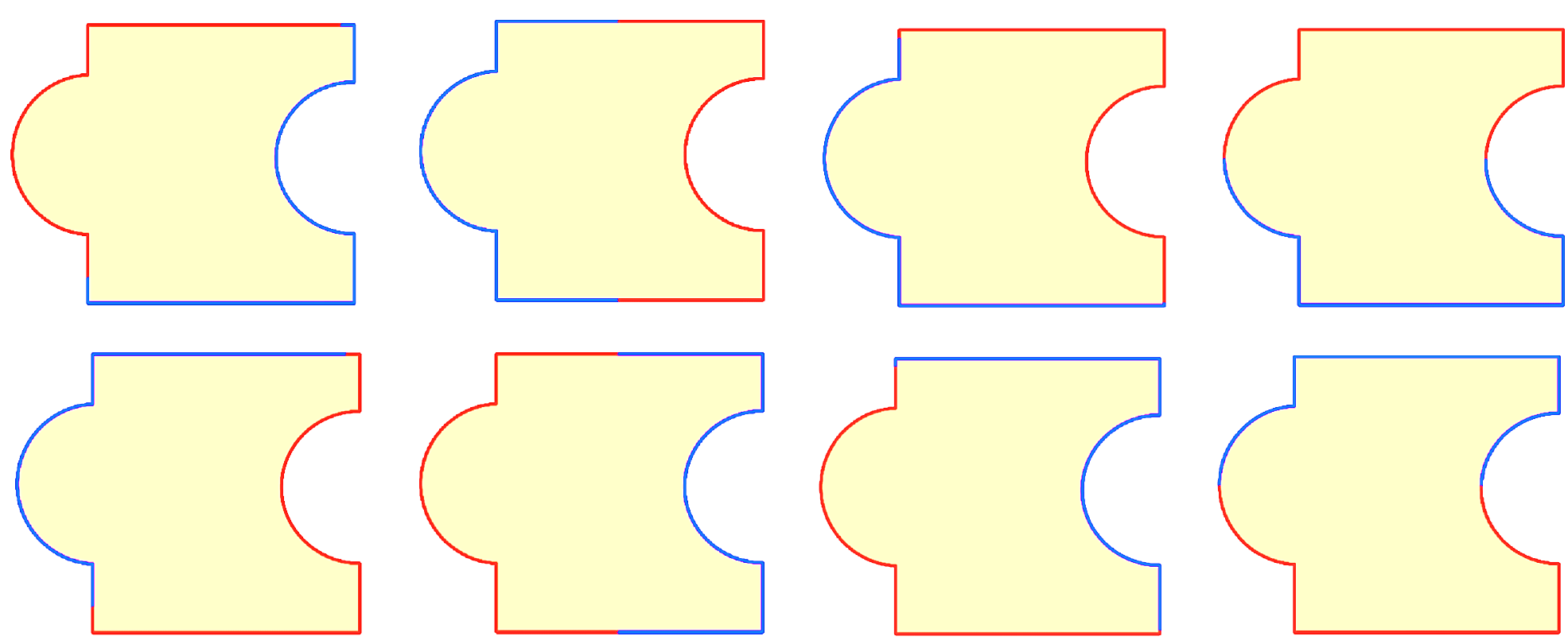}
  \caption[width=1\textwidth]{Example of boundary condition configurations used in the training set. Dirichlet boundaries are shown in red, and Neumann boundaries are shown in blue.}
\label{Fig:BCTypes}
\end{figure}

To train the hybrid CNN-MLP network for each MPS operator, we construct a dataset of boundary contributions from diverse 2D test cases with the following characteristics:

\begin{itemize}
    \item \textbf{Geometry:} Four complex geometries featuring various shapes, including convex and concave polygons as well as curved boundaries (Fig.~\ref{Fig:Geometries}).

    \item \textbf{Particle and kernel size:} Three particle diameters ($d_p = 0.01$,m, $0.02$,m, $0.03$,m) and five kernel radii ($r_e = 2.1d_p$, $2.6d_p$, $3.1d_p$, $3.6d_p$, $4.1d_p$) are considered.

    \item \textbf{Variable field:} For scalar fields, 62 predefined functions are used (summarized in Eq.~\eqref{Eqns:train}), capturing combinations of linear, polynomial, or periodic variations (with various frequencies). For vector fields, we define $\bm{\upphi}_i = (\phi_i, -\phi_i)$.

    \begin{equation}
    \begin{aligned}
        \phi(x, y) &= \pm f_1(2\alpha \pi x) \cdot f_2(2\beta \pi y),
        && \quad f_1, f_2 \in \{\sin, \cos\},\quad \alpha, \beta \in \{1,2\} \\
        \phi(x, y) &= \pm x^m y^n,
        && \quad m, n \in \{0, 1,2,3\}
    \end{aligned}
    \label{Eqns:train}
    \end{equation}
    
    \item \textbf{Boundary conditions:} Each geometry is simulated under eight boundary condition configurations, with half of the boundary length assigned a fixed value (Dirichlet condition) and the other half a zero-gradient (Neumann condition) (Fig.~\ref{Fig:BCTypes}).
\end{itemize}

For each MPS operation, this setup yields approximately 37,000 test cases and around 24 million data rows, based on the number of particles per case. However, most particles are unaffected by boundaries (i.e., they have no wall particles within their kernel support), which reduces training efficiency and accuracy. Therefore, these particles are filtered out, leaving a refined dataset of approximately 7.75 million rows. Of this, 80\% ($\approx 6.2$ millions) is used for training, 10\%  ($\approx 775{,}000$) for validation, and 10\%  for testing. The labels are generated using MPS with the ghost particle approach. 

\subsection{Test cases}

\begin{figure}[H]
    \centering
    \includegraphics[width=1\textwidth]{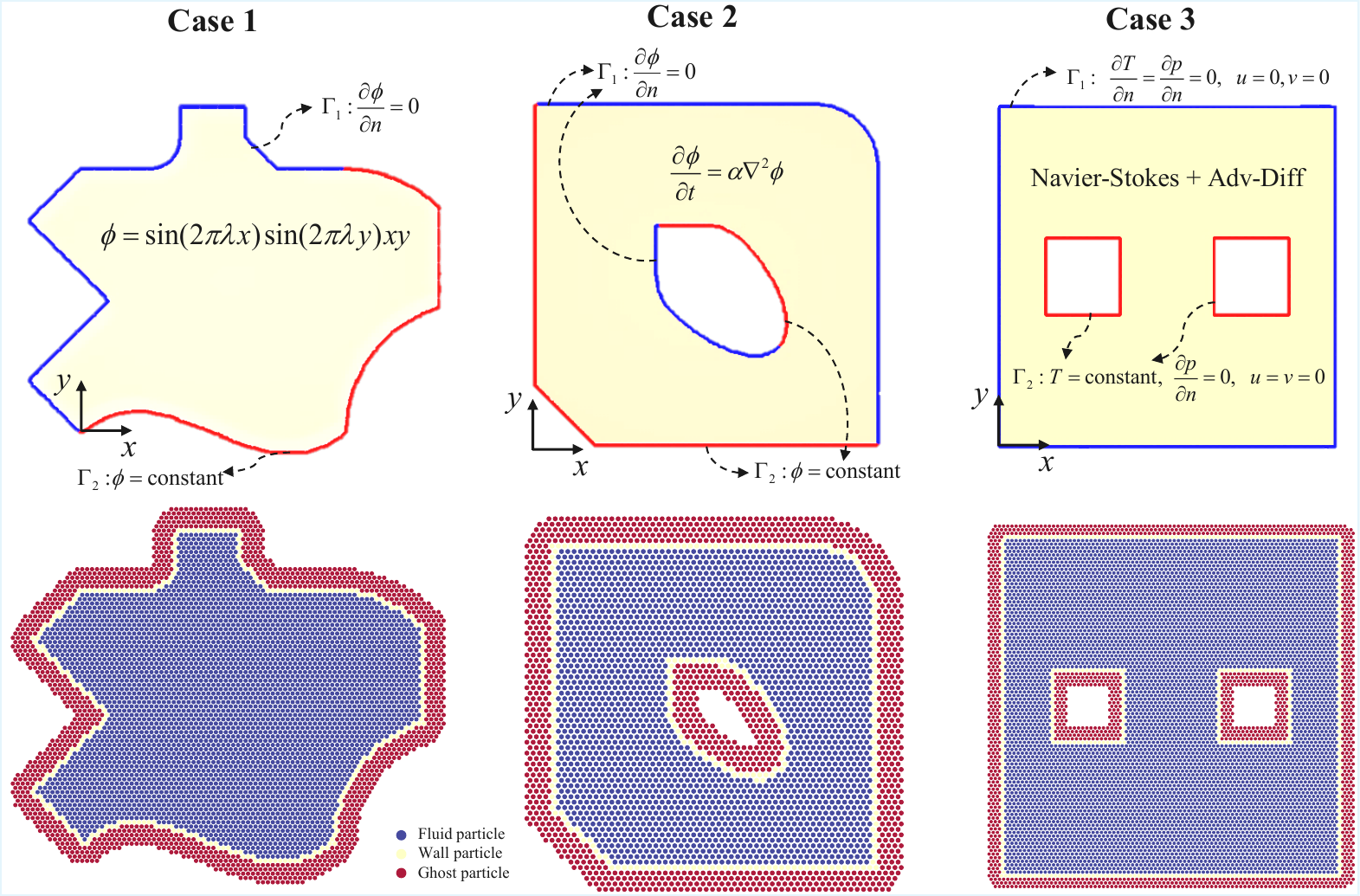}
    \caption[width=1\textwidth]{Geometry and boundary conditions (top row), and the particle representation and types (bottom row) for three test cases of this study.
}
    \label{Fig:cases}
\end{figure}

We evaluate the model's generalizability on three test cases of increasing complexity, all featuring unseen conditions. Each case has a geometry that differs from both the training data and the other test cases. A key distinction lies in how the field variable is defined: in \textbf{Case 1}, it is predefined using an analytical function; in \textbf{Case 2}, it is derived from the solution of an unsteady diffusion PDE; and in \textbf{Case 3}, it results from solving the unsteady Navier–Stokes equations coupled with advection–diffusion (representing flow and heat transfer). Figure~\ref{Fig:cases} illustrates the geometry, boundary conditions, and particle distribution for the test cases. Note that ghost particles are employed only in the ghost particle approach used to evaluate the developed model. Table \ref{tab:test_cases} summarizes characteristics of these three test cases. 

\textbf{Case 1 (Predefined Function)}: Figure~\ref{Fig:cases} shows the geometry, boundary conditions, and particle representation of this case.  The geometry includes complex convex and concave curves and corners, distinct from those in the training dataset. Boundary conditions are shown in are zero-gradient (Neumann) and fixed-value (Dirichlet) boundaries. The case uses a particle diameter of 0.02\,m and a kernel radius of $3.1d_p$. A new set of analytical functions, not included in the training set, is used to define the scalar field as:
\label{sec:case1}
\begin{equation}
\phi  = \sin(2\pi \lambda x){\mkern 1mu} \sin(2\pi \lambda y)xy, \quad 
\lambda \in \{ 2,3,4\}
\label{eqn:predef}
\end{equation}

The functions are periodic with variable spatial frequencies, including values higher than those present in the training data.  Both the geometry and the field functions in this case are more complex than those used in training. The trained hybrid CNN-MLP model predicts the boundary contributions to all MPS operators, including particle density, and scalar/vector gradients and Laplacians.

\textbf{Case 2 (Unsteady pure diffusion}):  Boundary conditions again include the Dirichlet and Neumann types. The case uses a particle diameter of 0.02\,m and a kernel radius of $3.1d_p$. In this case, the variable field is given by the solution of a PDE, i.e., 2D unsteady pure diffusion as:

\begin{equation}
    \frac{\partial \phi}{\partial t} = \alpha \nabla^2 \phi
\end{equation}
with diffusion coefficient $\alpha = 1\ \text{m}^2/\text{s}$. The only spatial derivative in this case is the scalar Laplacian, which is approximated using the MPS formulation. The trained model predicts the boundary contribution to the Laplacian operator without relying on ghost particles, and the results are compared with those obtained using the ghost particle method. Time integration is performed using the explicit Euler method with a time step $\Delta t = 0.001$s.

\textbf{Case 3 (Navier--Stokes with advection--diffusion)}: This test case involves coupled flow and heat transfer, governed by the unsteady incompressible Navier–Stokes equations with an advection–diffusion equation for temperature (see Section 2.1.2 for governing equations and MPS numerical solution). The geometry is shown in Figure~\ref{Fig:cases} and includes a square domain with two square cavities. The particle diameter is $d_p=0.0125 m$, and the kernel radius is set to $3.2d_p$.  

The temperature boundary conditions are defined with a fixed value of 1 (hot) on the left square cavity, a fixed value of -1 (cold) on the right square cavity, and zero-gradient (Neumann) conditions on all external boundaries. For this case, the density $\rho = 1 \ \mathrm{kg \, m^{-3}}$ , dynamic viscosity $\mu = 0.004 \ \mathrm{kg \, m^{-1} \, s^{-1}}$ , and the Rayleigh number $Ra=1.6 \times 10^6$ (see Section~\ref{app1}). This test case introduces additional complexity due to fluid motion, resulting in a dynamic particle distribution that is not encountered during training. The trained model predicts boundary contributions to the MPS operators in this unsteady, multi-physics scenario without relying on ghost particles. 

\begin{table}[H]
\centering
\caption{Summary of test case characteristics}
\label{tab:test_cases}
\resizebox{\textwidth}{!}{%
\begin{tabular}{|l|c|c|c|}
\hline
\textbf{Feature} & \textbf{Case 1} & \textbf{Case 2} & \textbf{Case 3} \\
\hline
Physics & None (predefined field) & Pure diffusion & Flow and heat transfer \\
Governing Equation & Analytical function: $\phi = \sin(2\pi \lambda x)  \sin(2\pi \lambda y)  x y$& PDE:
$\frac{\partial \phi}{\partial t} = \alpha \nabla^2 \phi$& PDEs:
(N.S. \& A.D.)\\
Particle Distribution & Static & Static & Dynamic \\
Boundary Conditions & Dirichlet, Neumann & Dirichlet, Neumann & Dirichlet, Neumann \\
Similarity to Training & Unseen geometry, similar physics & Unseen physics \& geometry & Unseen physics \& geometry \\
Particle Diameter ($d_p$) & 0.02 m & 0.02 m & 0.0125 m \\
Kernel Radius ($r$) & $3.1d_p$ & $3.1d_p$ & $3.2d_p$ \\
Examined operators & 
$n$, $\nabla \phi$, $\tilde\nabla \phi$, $\nabla \cdot \bm{\upphi}$, $\nabla^2 \phi$, $\nabla^2 \bm{\upphi}$ & 
$n$, $\nabla^2 \phi$& 
$n$, $\tilde\nabla p$, $\nabla \cdot \mathbf{u}$, $\nabla^2 \mathbf{u}$, $\nabla^2 T$ \\
\hline
\end{tabular}%
}
\end{table}

\section{Results and discussion}

\subsection{Training  evaluation}
As previously mentioned, the dataset consists of 7.75 million data points, of which 80\% were used for training, 10\% for validation, and the remaining 10\% for testing. The model was trained using a cross-validation trade-off to determine the optimal weights of the CNN–MLP (see ~\ref{app2}). To evaluate the accuracy of the trained hybrid CNN-MLP model, in addition to the Pearson correlation coefficient ($R$) discussed earlier, we employ two additional metrics: the normalized mean absolute error (NMAE) and the normalized root mean square error (NRMSE), defined as: 

\begin{equation}
\text{NMAE}(\%) =
\frac{1}{N} \sum_{i=1}^N\frac{| \hat{B}_i - B_i |}{\max(B) - \min(B)} \times 100,
\label{eq:aspe}
\end{equation}

\begin{equation}
\text{NRMSE}~(\%) = \frac{ \sqrt{ \frac{1}{N} \sum_{i=1}^{N} \left( \hat{B}_i - B_i \right)^2 } }{ \max(B) - \min(B) } \times 100
\end{equation}

where $\hat{B}$  and ${B}$ are predicted and true (from ghost particle method) boundary contributions as in equations \ref{Eq:B}, and  \ref{Eq:B_hat}, respectively.  The model's performance is evaluated for particle number density $n$ and five differential operators ($n$, \( \nabla\phi \), \( \tilde\nabla\phi \), \( \nabla \cdot \bm{\upphi} \), \( \nabla^2\phi \), and \( \nabla^2\bm{\upphi} \))  as summarized in Table~\ref{table:trainingErrors}. The results show consistently low errors and high accuracy across all datasets, indicating excellent model generalization, robustness, and no signs of overfitting. Slightly higher errors are observed for the second-order operators, particularly the vector Laplacian \( \nabla^2\bm{\upphi} \), likely due to their greater sensitivity to local variations and noise in the data.

\begin{table}[H]
\centering
\caption{Model Performance for training dataset}
\resizebox{\textwidth}{!}{
\begin{tabular}{llrrrrrrrrr}
\toprule
\multirow{2}{*}{\textbf{Case}} & \multirow{2}{*}{\textbf{Metrics}} & \multicolumn{9}{c}{\textbf{Targets}} \\
\cmidrule(l){3-11}
 &  & $n$ & $\nabla \phi_x$ & $\nabla \phi_y$  & $\tilde\nabla {\phi}_x$  & $\tilde\nabla {\phi}_y$ & $\nabla^2 \phi$ & $\nabla \cdot  \bm{\upphi}$  &  $\nabla^2  \bm{\upphi}_x$  & $\nabla^2  \bm{\upphi}_y$ \\
\midrule
\multirow{4}{*}{Training}
 & R& 1.000 & 0.984 & 0.989 & 1.000 & 1.000 & 0.980 & 0.997 & 0.999 & 0.999 \\
 & NMAE (\%)& 0.253 & 0.369 & 0.338 & 0.067 & 0.075 & 0.385 & 0.170 & 0.074 & 0.079 \\
 & NRMSE (\%)  & 0.390 & 0.965 & 0.767 & 0.176 & 0.171 & 1.611 & 0.500 & 0.265 & 0.280 \\
\midrule
\multirow{4}{*}{Validation}
  & R& 1.000 & 0.979 & 0.987 & 1.000 & 1.000 & 0.972 & 0.996 & 0.999 & 0.999 \\
  & NMAE (\%)& 0.254 & 0.390 & 0.353 & 0.070 & 0.078 & 0.429 & 0.186 & 0.078 & 0.082 \\
 & NRMSE (\%)  & 0.390 & 1.099 & 0.852 & 0.189 & 0.183 & 1.914 & 0.584 & 0.272 & 0.273 \\
 \midrule
 \multirow{4}{*}{Testing}
& R& 1.000 & 0.979 & 0.987 & 1.000 & 1.000 & 0.972 & 0.996 & 0.998 &0.998\\ 
& NMSE(\%)& 0.253 & 0.389 & 0.353 & 0.071 & 0.078 & 0.426 & 0.185 & 0.078 &0.082 \\
 & NRMSE (\%)  & 0.391 & 1.109 & 0.846 & 0.194 & 0.181 & 1.888 & 0.579 & 0.278 &0.296 \\
 \midrule
\end{tabular}}
\label{table:trainingErrors}
\end{table}

\subsection{Generalizability beyond training conditions}
\subsubsection{Case 1: Predefined Function}

\begin{figure}[H]
  \centering
  \includegraphics[width=.9\textwidth]{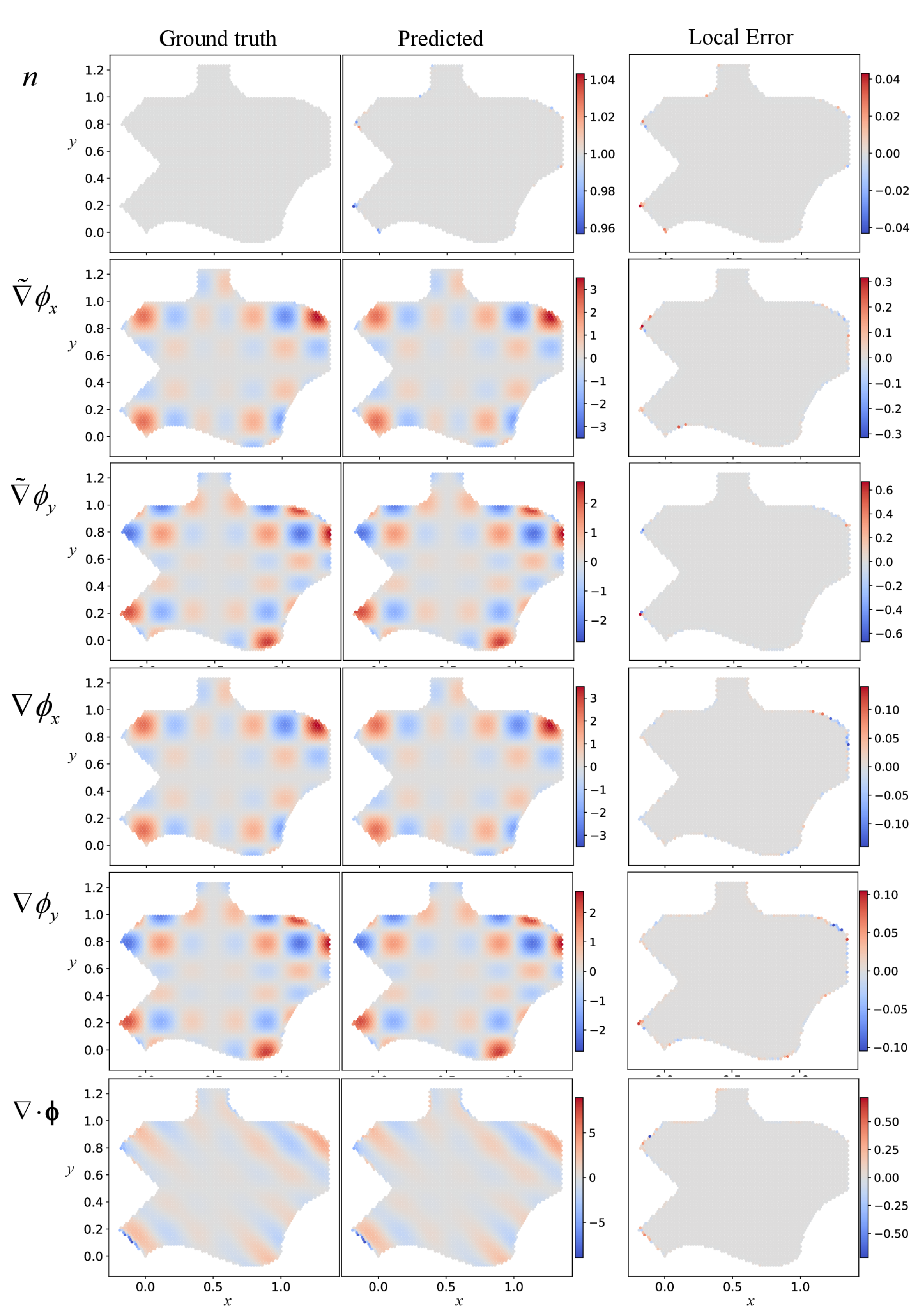}
  \caption[width=0.7\textwidth]{MPS-predicted fields with ML-based boundary treatment ($\hat C_i$), 
ground-truth MPS fields with ghost-particle boundary treatment ($C_i$), 
and local error ($C_i - \hat C_i$) for particle number density ($n$) and first-order derivatives ($\nabla \phi$, $\tilde \nabla \phi$, and $\nabla \cdot \bm{\upphi}$) in Case 1 ($\lambda = 2$)}
  \label{Fig:predefined_n0_and_First_derivatives}
\end{figure}

\begin{figure}[H]
  \centering
  \includegraphics[width=.9\textwidth]{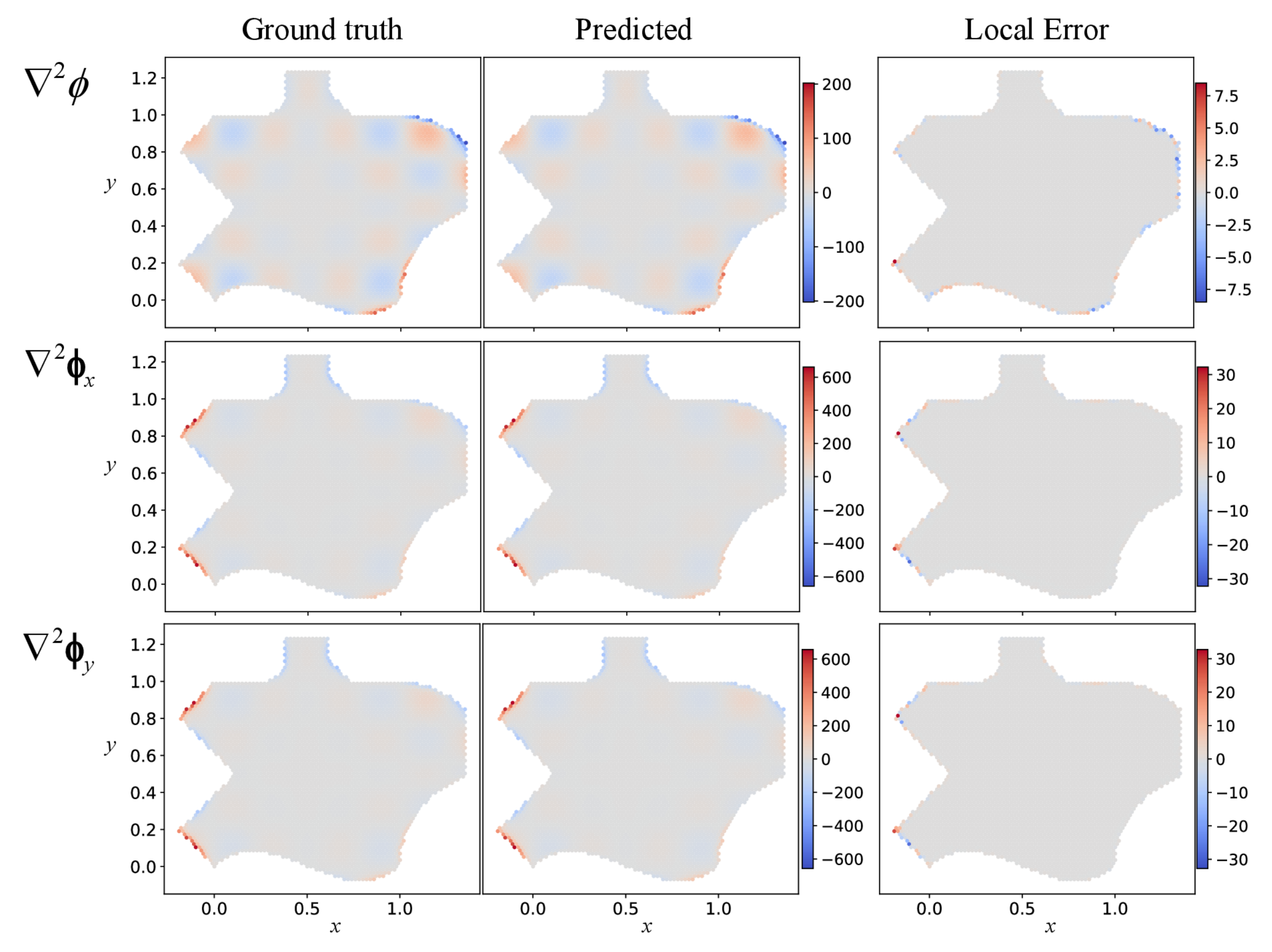}
  \caption[width=0.7\textwidth]{MPS-predicted fields with ML-based boundary treatment ($\hat C_i$), 
ground-truth MPS fields with ghost-particle boundary treatment ($C_i$), 
and local error ($C_i - \hat C_i$) for second-order derivatives ($\nabla^2 \phi$, $\nabla^2 \bm{\upphi}_x$, and $\nabla^2 \bm{\upphi}_y$) in Case 1 ($\lambda = 2$)}
  \label{Fig:predefined_Second_derivatives}
\end{figure}

Figures~\ref{Fig:predefined_n0_and_First_derivatives} and~\ref{Fig:predefined_Second_derivatives} present the MPS calculated particle number density, first-order derivatives, and second-order derivatives, where the boundary contributions are predicted by the developed ML model, alongside the corresponding ground truth values computed using the ghost particle method. These figures also display the difference between the predicted and reference fields, enabling a visual assessment of the model’s accuracy.

Overall, excellent agreement is observed between the ML predictions and the ground truth for all operators. The predicted fields closely match the true values, demonstrating the model’s capability to generalize to new, unseen test cases. Minor discrepancies are observed near a few sharp boundary corners, and this behavior appears consistently across all operators, including the particle number density. Since the boundary contribution of the particle number density depends only on geometry (with no dependence on field variables), these deviations can be attributed to geometrical complexities not captured during training.

\begin{figure}[H]
  \centering
  \includegraphics[width=0.75\textwidth]{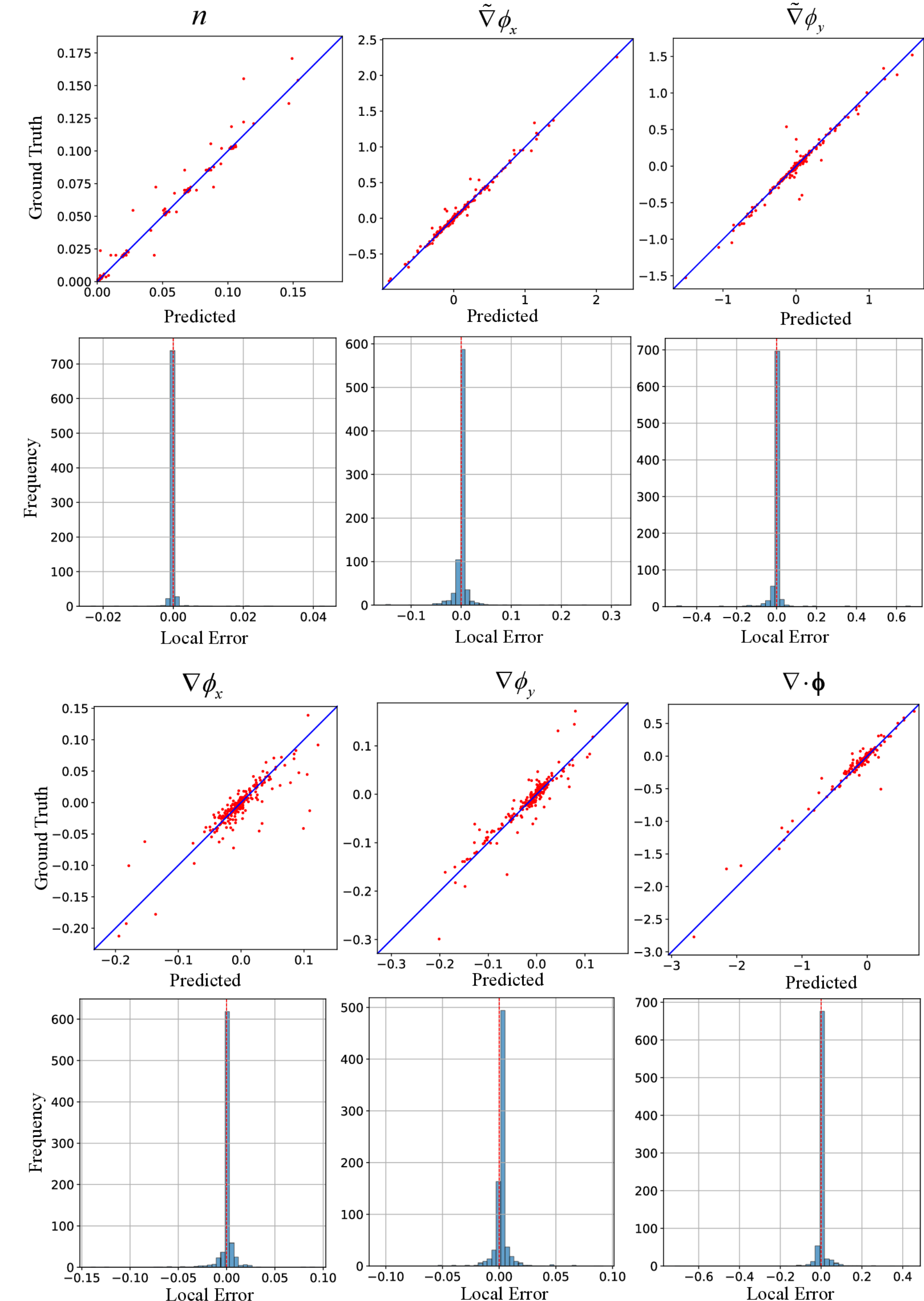}
  \caption[width=0.7\textwidth]{Parity plots (top row) and error histograms (bottom row) comparing predicted boundary contributions $\hat{B}_i$ with ground truth values ${B}_i$ for  particle number density ($n$) and first-order derivatives ($\nabla \phi$, $\tilde \nabla \phi$, and $\nabla \cdot \bm{\upphi}$)  in Case 1 ($\lambda = 2$) }
  \label{Fig:predefined_n0_and_First_derivatives_for_Histogram_and_errors}
\end{figure}

\begin{figure}[H]
  \centering
  \includegraphics[width=0.75\textwidth]{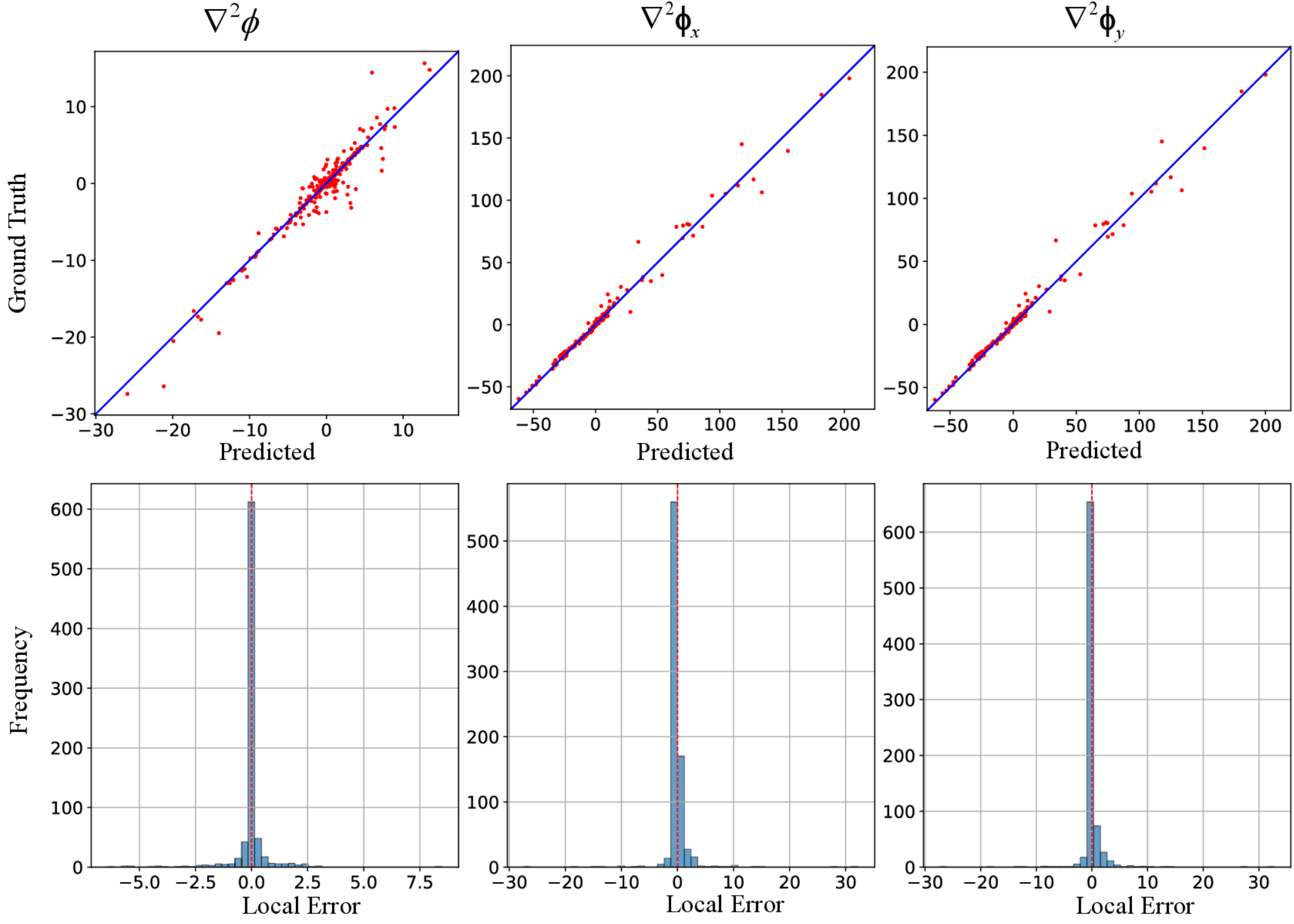}
\caption[width=0.7\textwidth]{Parity plots (top row) and error histograms (bottom row) comparing predicted boundary contributions $\hat{B}_i$ with ground truth values ${B}_i$ for second-order derivatives ($\nabla^2 \phi$, $\nabla^2  \bm{\upphi}_x$, and $\nabla^2 \bm{\upphi}_y$) in Case 1 ($\lambda = 2$).}
  \label{Fig:predefined_Second_derivatives_for_Histogram_and_errors}
\end{figure}

To further quantify the model's performance, Figures~\ref{Fig:predefined_n0_and_First_derivatives_for_Histogram_and_errors} and~\ref{Fig:predefined_Second_derivatives_for_Histogram_and_errors} show the parity plots (predicted vs. ground truth) and corresponding error histograms for each predicted quantity. The parity plots show a strong alignment along the ideal \( R=1 \) line, indicating high accuracy of the ML predictions. The error histograms show that most error values are tightly clustered around zero, confirming the model's reliability and precision. 

To evaluate the model’s generalization capacity, Table~\ref{table:predefinedErrors} compares the correlation coefficient \( R \), NMSE, and NRMSE of the ML predictions for different spatial frequency parameters \( \lambda \), including values beyond the training range (up to three times higher). Additionally, a case with artificially introduced non-uniformity in the particle distribution is considered, achieved by randomly perturbing particle positions using Gaussian noise with a standard deviation of \( \pm 2\% d_p \). This aims to account for the physical and numerical fluctuations that commonly occur during flow simulations in particle methods.

Among the differential operators evaluated, the ML model consistently shows the highest prediction accuracy for the Laplacian of vector components (\( \nabla^2 \bm{\upphi}_x \), \( \nabla^2 \bm{\upphi}_y \)) and the divergence operator (\( \nabla \cdot \bm{\upphi} \)). The scalar Laplacian (\( \nabla^2 \phi \)) is also predicted with good accuracy, exhibiting slightly higher errors than the vector components but remaining robust across all conditions. The model’s performance is slightly less for the gradient components (\( \nabla \phi_x \), \( \nabla \phi_y \)). Prediction of the particle number density \( n \) remains consistently accurate across all cases.

The results demonstrate that the developed model generalizes well across a range of spatial frequencies, maintaining consistently high correlation coefficients and NMSE and NRMSE  for most differential operators. The model also shows strong robustness to moderate particle distribution non-uniformity. Prediction accuracy remains reliable for both first- and second-order derivatives, as well as for the particle number density. Overall, only a modest decline in performance is observed as spatial frequency and non-uniformity increase, confirming the model’s robustness and generalization capacity for practical applications.

\begin{table}[]
\centering
\caption{Model performance for test case 1 for various particle distributions and frequency coefficients ($\lambda$)}
\resizebox{\textwidth}{!}{
\begin{tabular}{llrrrrrrrrr}
\toprule
\multirow{2}{*}{\textbf{Condition}} & \multirow{2}{*}{\textbf{Metrics}} & \multicolumn{9}{c}{\textbf{Targets}} \\
\cmidrule(l){3-11}
 &  & $n$ & $\nabla \phi_x$ & $\nabla \phi_y$  & $\tilde\nabla {\phi}_x$  & $\tilde\nabla {\phi}_y$ & $\nabla^2 \phi$ & $\nabla \cdot  \bm{\upphi}$  &  $\nabla^2  \bm{\upphi}_x$  & $\nabla^2  \bm{\upphi}_y$ \\
\midrule
\multirow{4}{*}{$\lambda=2$, Uniform}
  & R& 0.996 & 0.860 & 0.899 & 0.993 & 0.985 & 0.967 & 0.976 & 0.991 & 0.991 \\
 & NMAE(\%)& 0.350 & 1.146 & 1.010 & 0.220 & 0.259 & 0.653 & 0.321 & 0.243 & 0.244 \\
 & NRMSE (\%)  & 1.874 & 3.141 & 2.681 & 0.765 & 1.065 & 1.909 & 1.243 & 0.959 & 0.955 \\
\midrule
\multirow{4}{*}{$\lambda=3$, Uniform}
 & R& 0.996 & 0.895 & 0.923 & 0.974 & 0.983 & 0.947 & 0.987 & 0.994 & 0.994 \\ 
 & NMAE(\%)& 0.350 & 1.070 & 1.047 & 0.471 & 0.350 & 0.961 & 0.282 & 0.221 & 0.220 \\
 & NRMSE (\%)  & 1.874 & 3.160 & 2.941 & 1.842 & 1.269 & 2.826 & 0.812 & 0.762 & 0.758 \\
\midrule
\multirow{4}{*}{$\lambda=4$, Uniform}
 & R& 0.996 & 0.882 & 0.908 & 0.980 & 0.979 & 0.956 & 0.965 & 0.983 & 0.982 \\
 & NMAE(\%)& 0.350 & 1.186 & 0.979 & 0.481 & 0.458 & 0.578 & 0.504 & 0.302 & 0.301 \\
 & NRMSE (\%)  & 1.874 & 2.965 & 2.581 & 1.468 & 1.409 & 1.723 & 1.675 & 1.126 & 1.135 \\
\midrule
\multirow{4}{*}{$\lambda=2$, Nonuniform}
  & R& 0.996 & 0.890 & 0.918 & 0.978 & 0.973 & 0.950 & 0.980 & 0.988 & 0.988 \\
 & NMAE(\%)& 0.350 & 1.004 & 0.896 & 0.396 & 0.396 & 0.752 & 0.313 & 0.276 & 0.278 \\
 & NRMSE (\%)  & 1.874 & 2.831 & 2.470 & 1.394 & 1.515 & 2.318 & 1.123 & 1.111 & 1.108 \\
\bottomrule
\end{tabular}}
\label{table:predefinedErrors}
\end{table}

\subsubsection{Case 2: Unsteady pure diffusion}

\begin{figure}[H]
  \centering
  \includegraphics[width=0.75\textwidth]{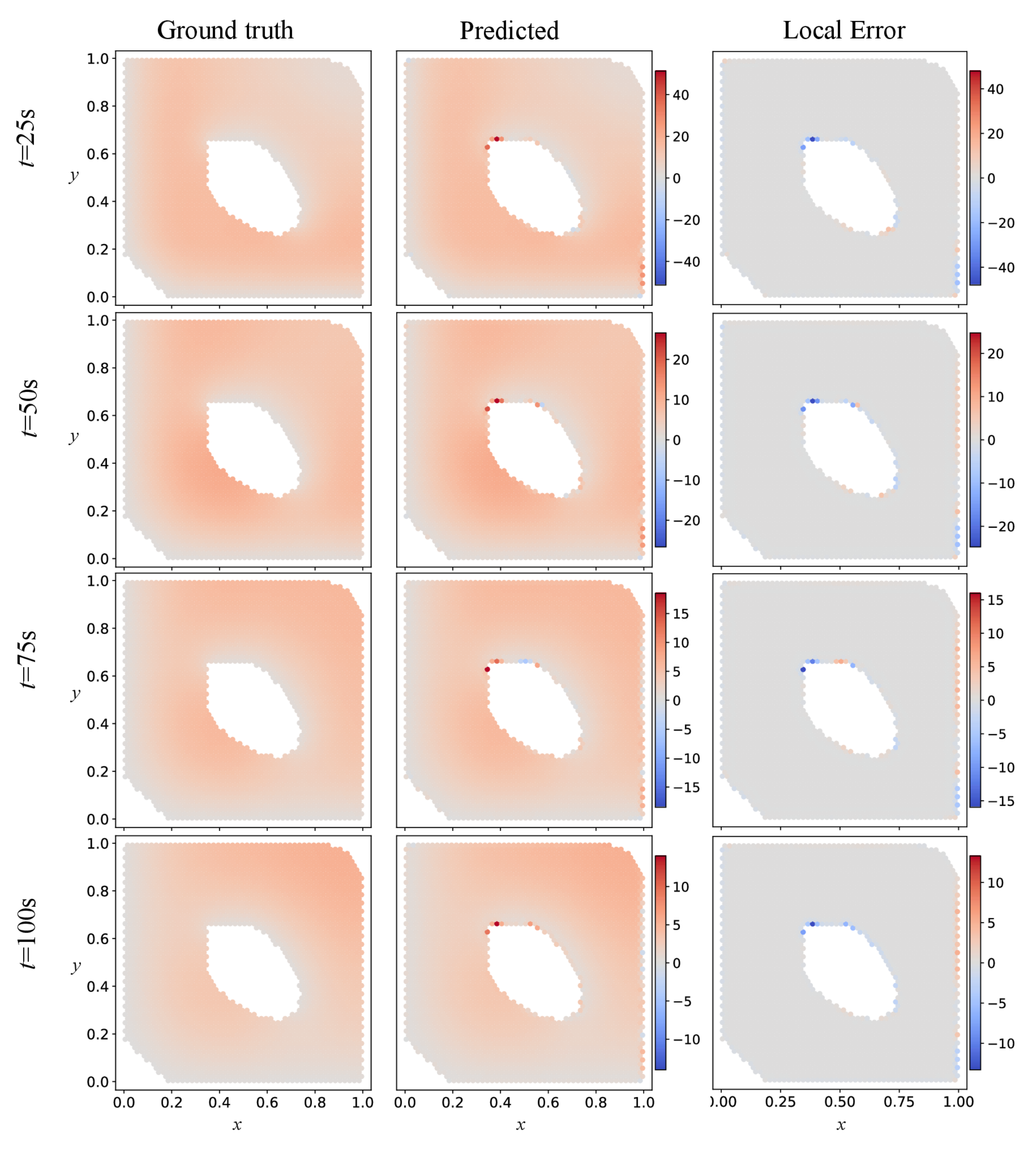}
  \caption[width=0.7\textwidth]{MPS-predicted fields with ML-based boundary treatment ($\hat C_i$), 
ground-truth MPS fields with ghost-particle boundary treatment ($C_i$), 
and local error ($C_i - \hat C_i$) for the laplacian of the scalar field ($\nabla^2 \phi$ )  in Case 2 }
  \label{Fig:pureDiff_laplacianScalar}
\end{figure}

\begin{figure}[H]
  \centering
  \includegraphics[width=1\textwidth]{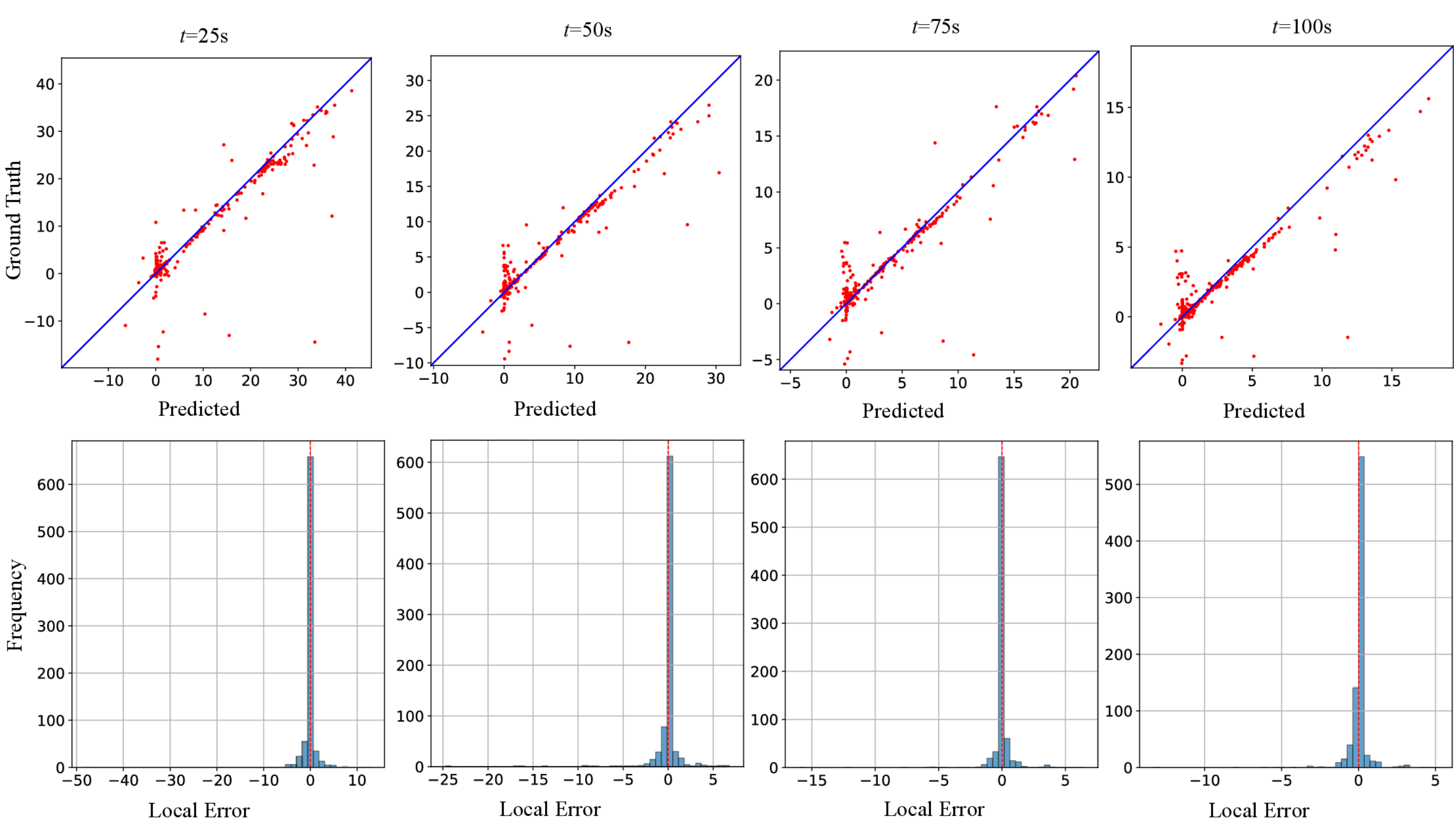}
  \caption[width=0.7\textwidth]{Parity plots (top row) and error histograms (bottom row) comparing predicted boundary contributions $\hat{B}_i$ with ground truth values ${B}_i$ for  particle number density ($n$) and laplacian of the scalar field ($\nabla^2 \phi$, )  in Case 2 .}
  \label{Fig:pureDiff_laplacianScalar_for_Histogram_and_errors}
\end{figure}

Figure~\ref{Fig:pureDiff_laplacianScalar} illustartes a comparison of  MPS results for the only spatial derivative in this case, $\nabla^2 \phi$, where the boundary contributions are provided by the ghost particle method (ground truth) and the predictions from the developed ML model. The difference between the two methods has also been provided. Overall, the ML model demonstrates excellent agreement with the ground truth at all simulation times. Aside from minor discrepancies near sharp corners, the predicted fields closely match the reference values. This highlights the model’s strong generalization capability to handle cases with non-predefined, spatially variable fields and an unseen domain featuring a central hole. The parity plots and error histograms in Figure~\ref{Fig:pureDiff_laplacianScalar_for_Histogram_and_errors} further support these findings, showing a high correlation between predicted and true boundary contributions, with most errors tightly clustered around zero. Table~\ref{table:pureDiffErrors} confirms that the model consistently maintains low error levels in this test, with performance metrics even better than those observed in Case 1.

\begin{table}[h]
\centering
\caption{Performance metrics for Case 2.}
\begin{tabular}{lrrrrr}
\toprule
 & \multicolumn{4}{c}{$\nabla^2 \phi$} & $n$ \\
\cmidrule(lr){2-5} \cmidrule(lr){6-6}
 \textbf{Metrics}& $t=25$ & $t=50$ & $t=75$ & $t=100$ &  \\
\midrule
R          & 0.944 & 0.946 & 0.942 & 0.938 & 0.999 \\
NMAE (\%)   & 1.220 & 1.284 & 1.160 & 1.403 & 0.394 \\
NRMSE (\%)  & 4.894 & 4.673 & 4.212 & 4.600 & 1.345 \\
\bottomrule
\end{tabular}

\label{table:pureDiffErrors}
\end{table}

\subsubsection{Case 3: Unsteady Navier-Stokes with advection-diffusion}

\begin{figure}[H]
  \centering
  \includegraphics[width=\textwidth]{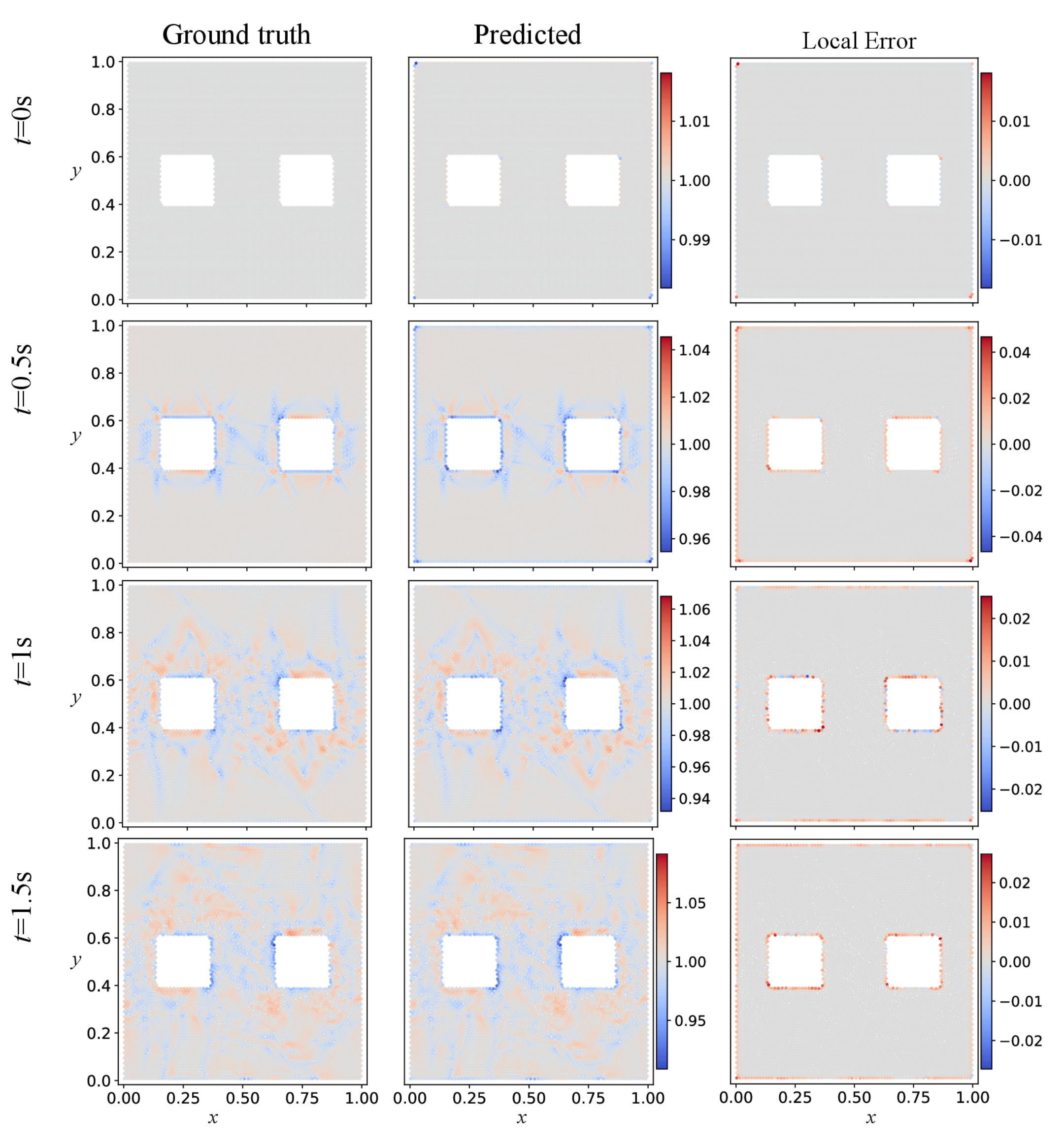}
  \caption[width=0.7\textwidth]{MPS-predicted fields with ML-based boundary treatment ($\hat C_i$), 
ground-truth MPS fields with ghost-particle boundary treatment ($C_i$), 
and local error ($C_i - \hat C_i$) for particle number density ($n$) in Case 3}
  \label{Fig:adv-diff_n0}
\end{figure}

\begin{figure}[H]
  \centering
  \includegraphics[width=1\textwidth]{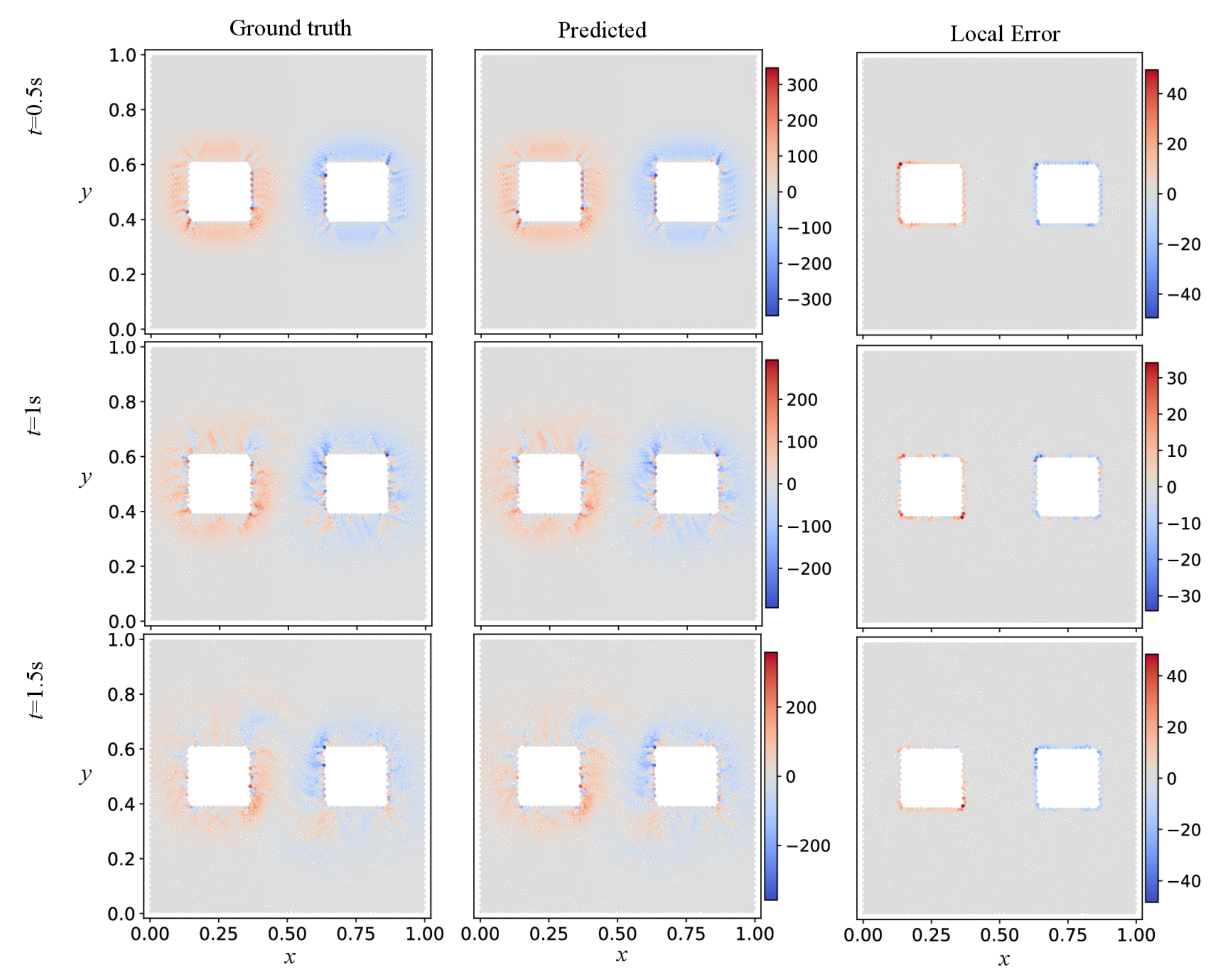}
  \caption[width=1\textwidth]{MPS-predicted fields with ML-based boundary treatment ($\hat C_i$), 
ground-truth MPS fields with ghost-particle boundary treatment ($C_i$), 
and local error ($C_i - \hat C_i$) for laplacian of the temprature field ($\nabla^2 T$ ) in Case 3}
  \label{Fig:adv-diff laplacianTemprature}
\end{figure}

\begin{figure}[H]
  \centering
  \includegraphics[width=.9\textwidth]{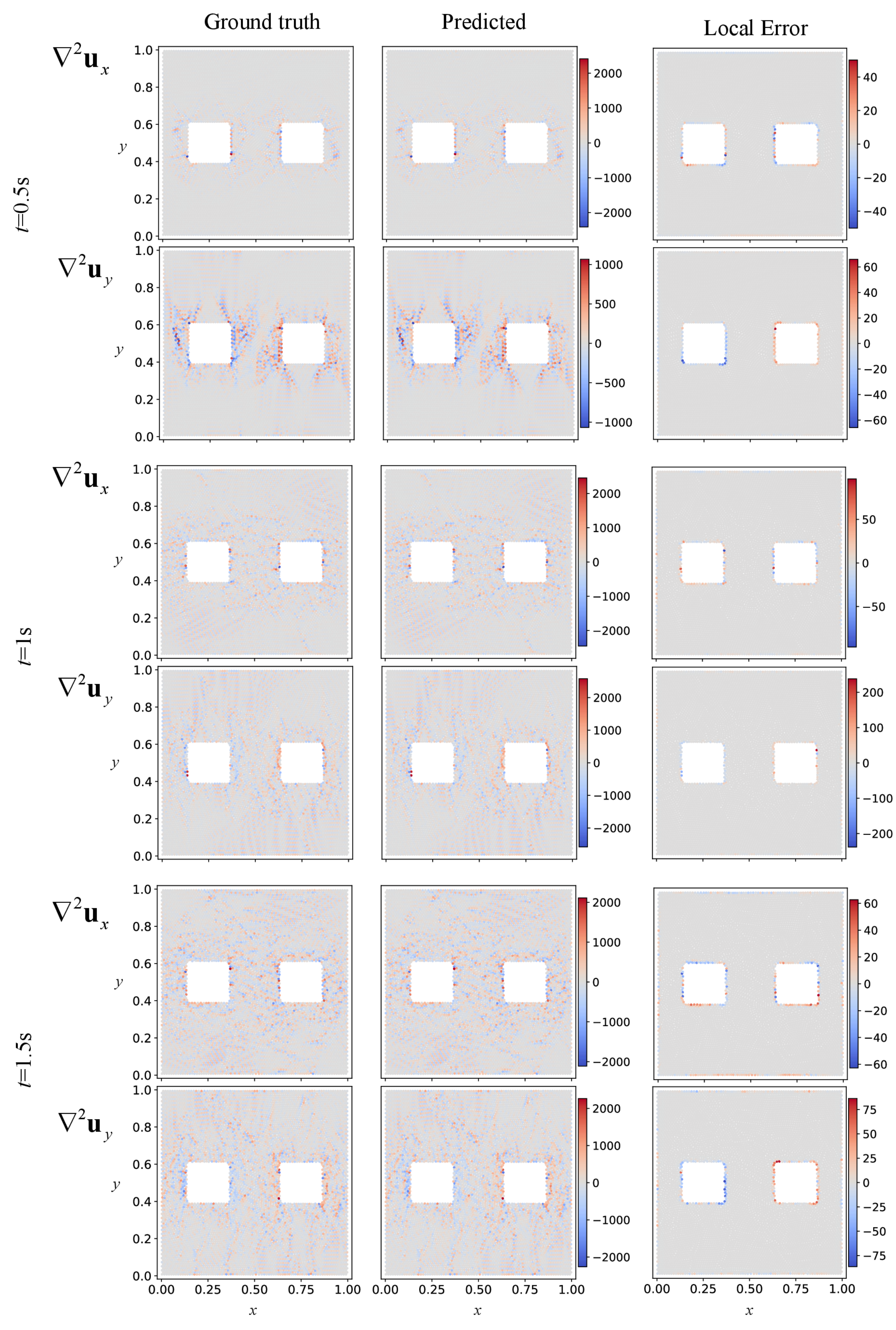}
  \caption[width=0.7\textwidth]{MPS-predicted fields with ML-based boundary treatment ($\hat C_i$), 
ground-truth MPS fields with ghost-particle boundary treatment ($C_i$), 
and local error ($C_i - \hat C_i$) for the laplacian of velocity ($\nabla^2 \boldsymbol{u}_x$, $\nabla^2 \boldsymbol{u}_y$) in Case 3}
  \label{Fig:adv-diff laplacianVelocity}
\end{figure}

\begin{figure}[H]
  \centering
  \includegraphics[width=\textwidth]{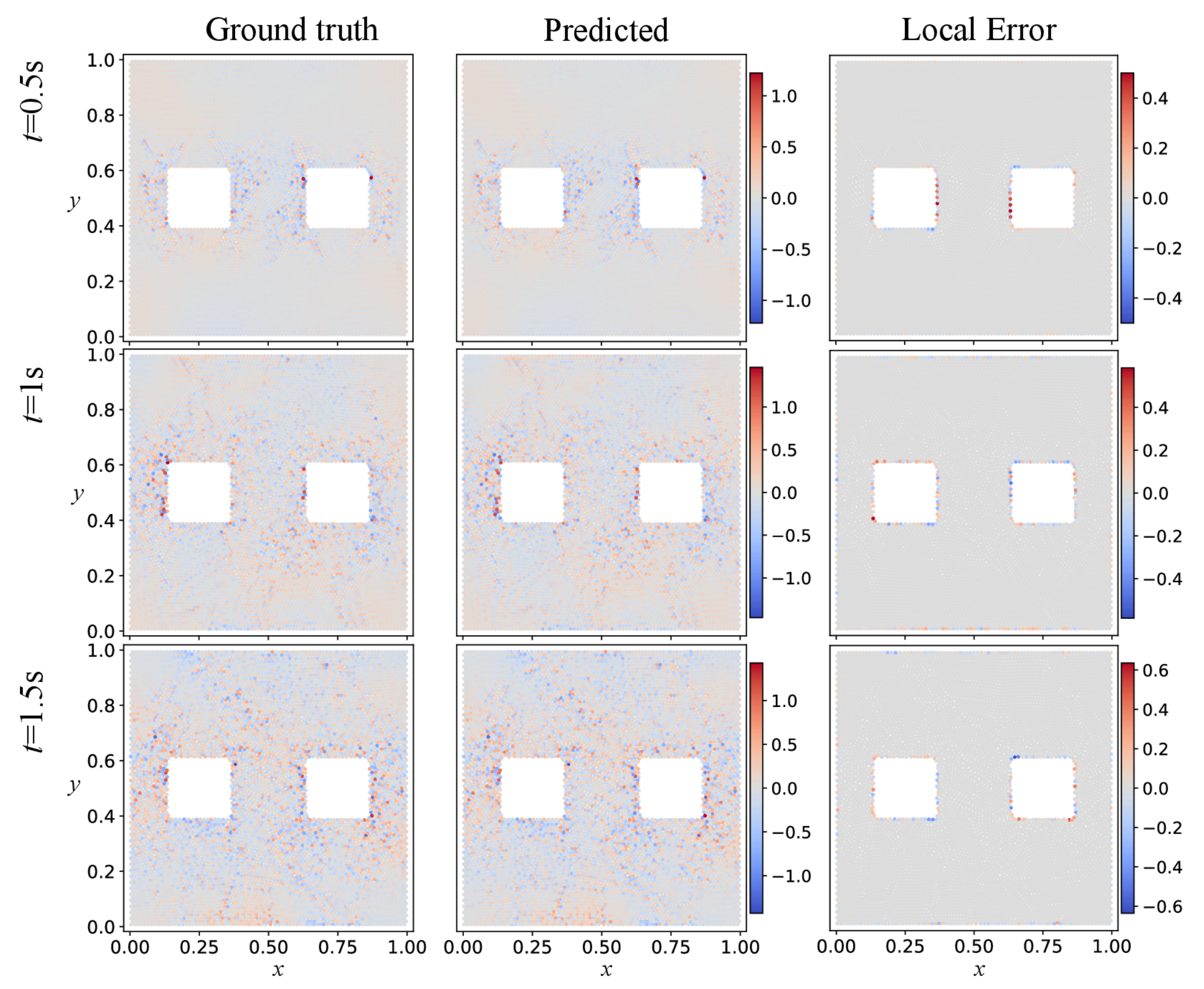}
  \caption[width=0.7\textwidth]{MPS-predicted fields with ML-based boundary treatment ($\hat C_i$), 
ground-truth MPS fields with ghost-particle boundary treatment ($C_i$), 
and local error ($C_i - \hat C_i$) for particle number density ($n$) and first-order derivatives ($\nabla \cdot \boldsymbol{u}$) in Case 3}
  \label{Fig:adv-diff divergenceVelocity}
\end{figure}

\begin{figure}[H]
  \centering
  \includegraphics[width=.9\textwidth]{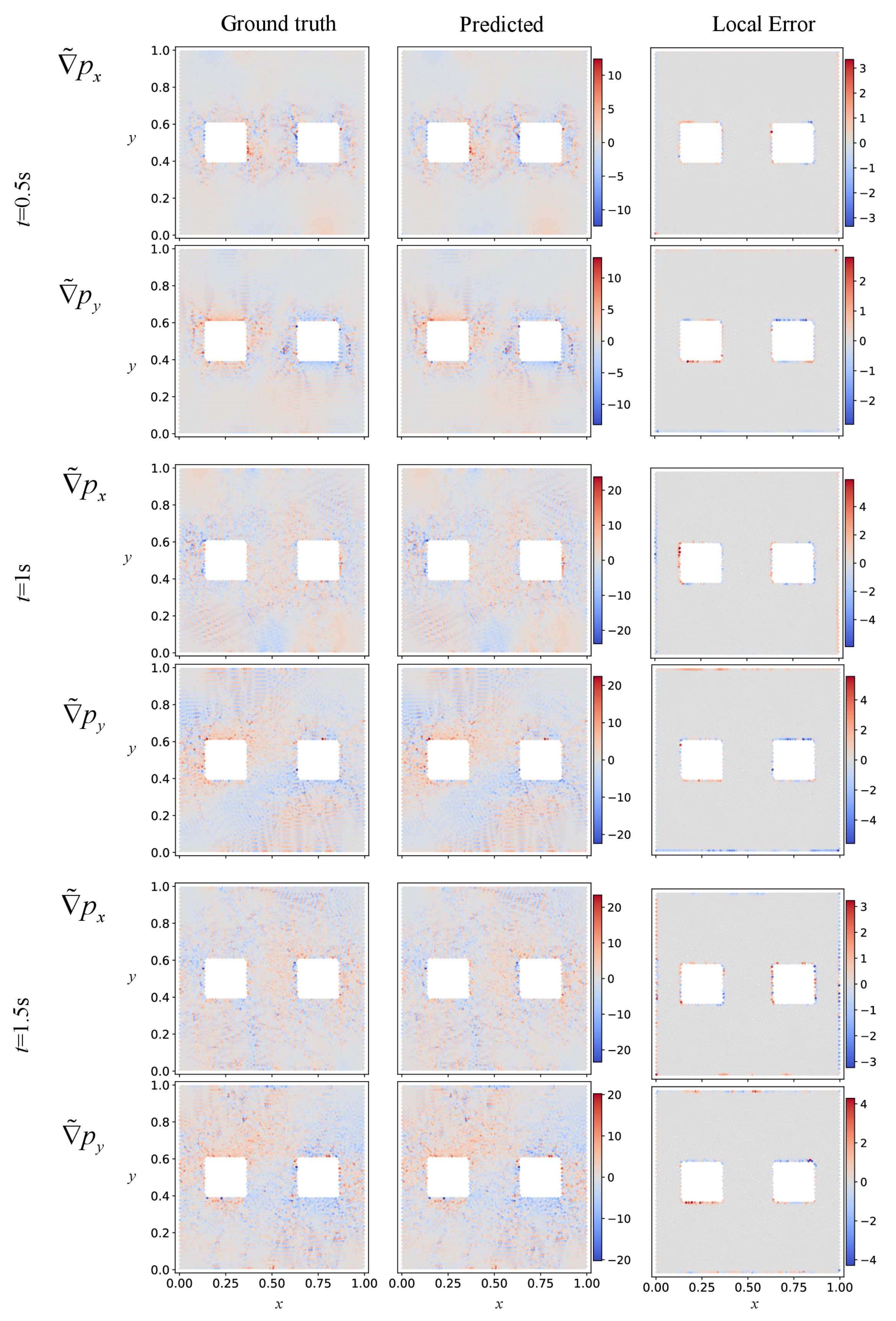}
  \caption[width=0.7\textwidth]{MPS-predicted fields with ML-based boundary treatment ($\hat C_i$), 
ground-truth MPS fields with ghost-particle boundary treatment ($C_i$), 
and local error ($C_i - \hat C_i$) for pressure gradient ($\nabla p_x$, $\nabla p_y$) in Case 3}
  \label{Fig:adv-diff gradientPressure}
\end{figure}

This test case introduces new generalization challenges due to its dynamic nature, characterized by spatial and temporal variability in both the fields and particle distributions, conditions not encountered during training. Figures \ref{Fig:adv-diff_n0}, \ref{Fig:adv-diff laplacianTemprature}, \ref{Fig:adv-diff laplacianVelocity}, \ref{Fig:adv-diff divergenceVelocity}, \ref{Fig:adv-diff gradientPressure} compare MPS results for different spatial derivatives in case 3, where boundary contributions are provided by the ghost particle method (ground truth) and the ML model predictions. Across all time steps, the predicted fields closely match the ground truth, with errors remaining relatively low and mostly confined to regions near the boundaries of the internal holes, particularly around sharp corners.

The error metrics in Table~\ref{table:adv-diffErrors} show that the trained hybrid CNN–MLP model generalizes well to this unseen scenario. Compared with the training dataset performance (Table~\ref{table:trainingErrors}), the correlation coefficients remain high ($R>0.94$ for all derivatives), demonstrating that the model continues to capture the underlying relationships between boundary values and spatial derivatives. Although normalized errors increase relative to the training case, particularly for second-order terms such as $\nabla^2 T$ and $\nabla^2 \boldsymbol{u}$, the overall accuracy remains satisfactory, with NRMSE values typically below 4\%. This suggests that the model is reliable for predicting both primary and higher-order derivative quantities, even though the latter are more sensitive to generalization errors due to their dependence on local variations.

\begin{figure}[H]
  \centering
  \includegraphics[width=1\textwidth]{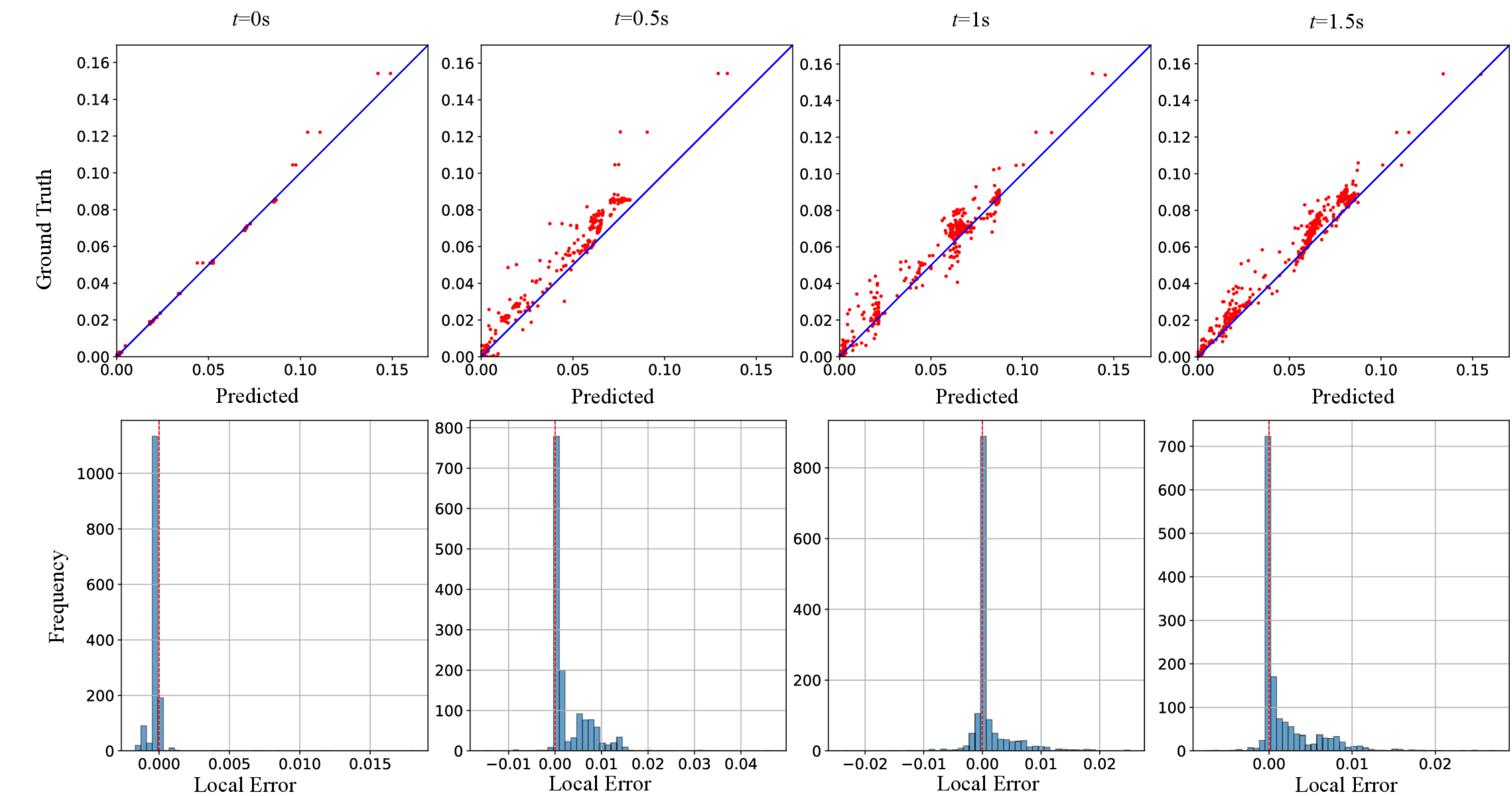}
  \caption[width=0.7\textwidth]{Parity plots (top row) and error histograms (bottom row) comparing predicted boundary contributions $\hat{B}_i$ with ground truth values ${B}_i$ for particle number density ($n$) in Case 3}
  \label{Fig:adv-diff_n0_for_Histogram_and_errors}
\end{figure}

\begin{figure}[H]
  \centering
  \includegraphics[width=.9\textwidth]{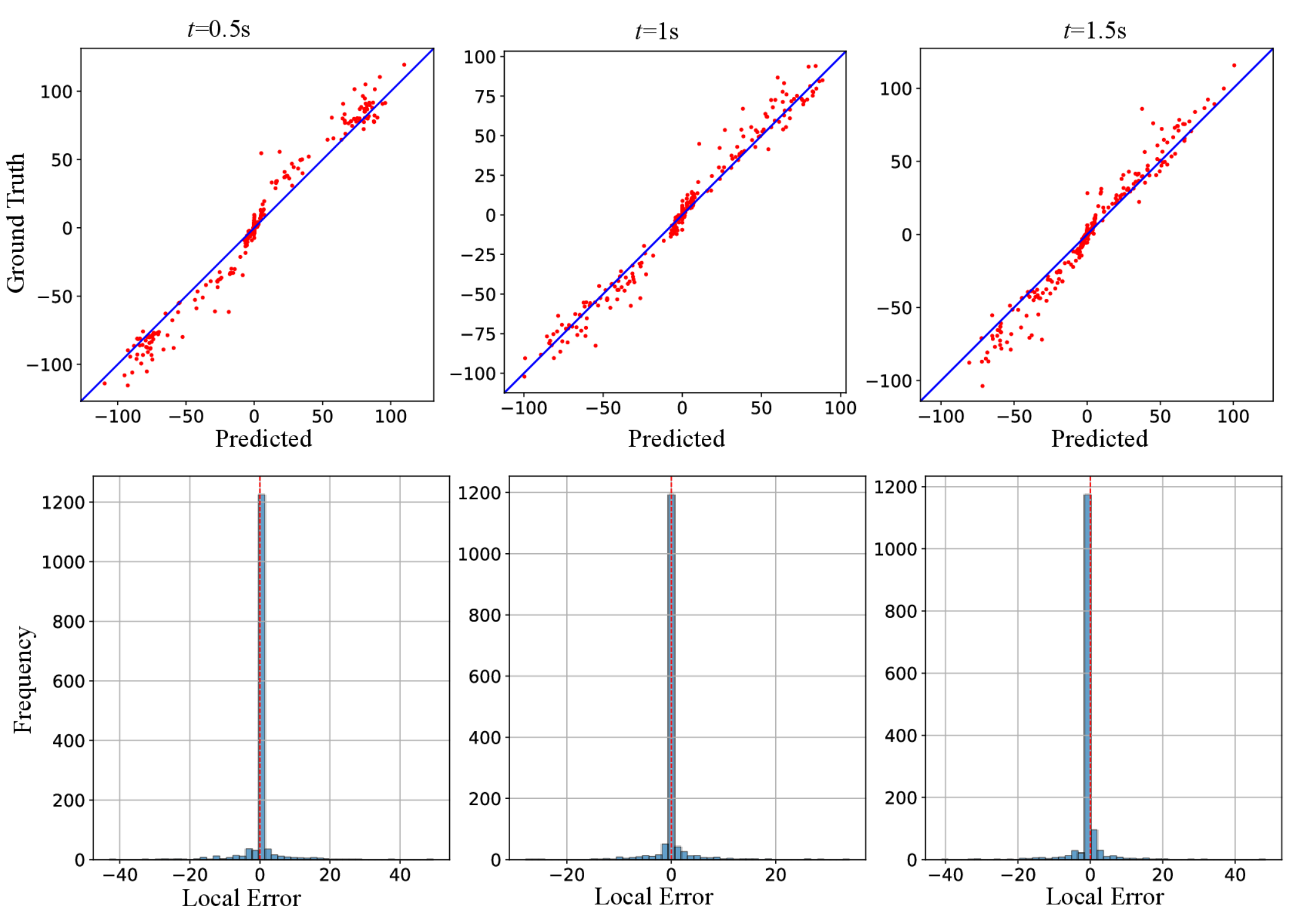}
  \caption[width=0.7\textwidth]{Parity plots (top row) and error histograms (bottom row) comparing predicted boundary contributions $\hat{B}_i$ with ground truth values ${B}_i$ for Laplacian of temperature ($\nabla^2 T$) in Case 3}
  \label{Fig:adv-diff laplacianTemprature for Histogram and errors}
\end{figure}

\begin{figure}[H]
  \centering
  \includegraphics[width=.9\textwidth]{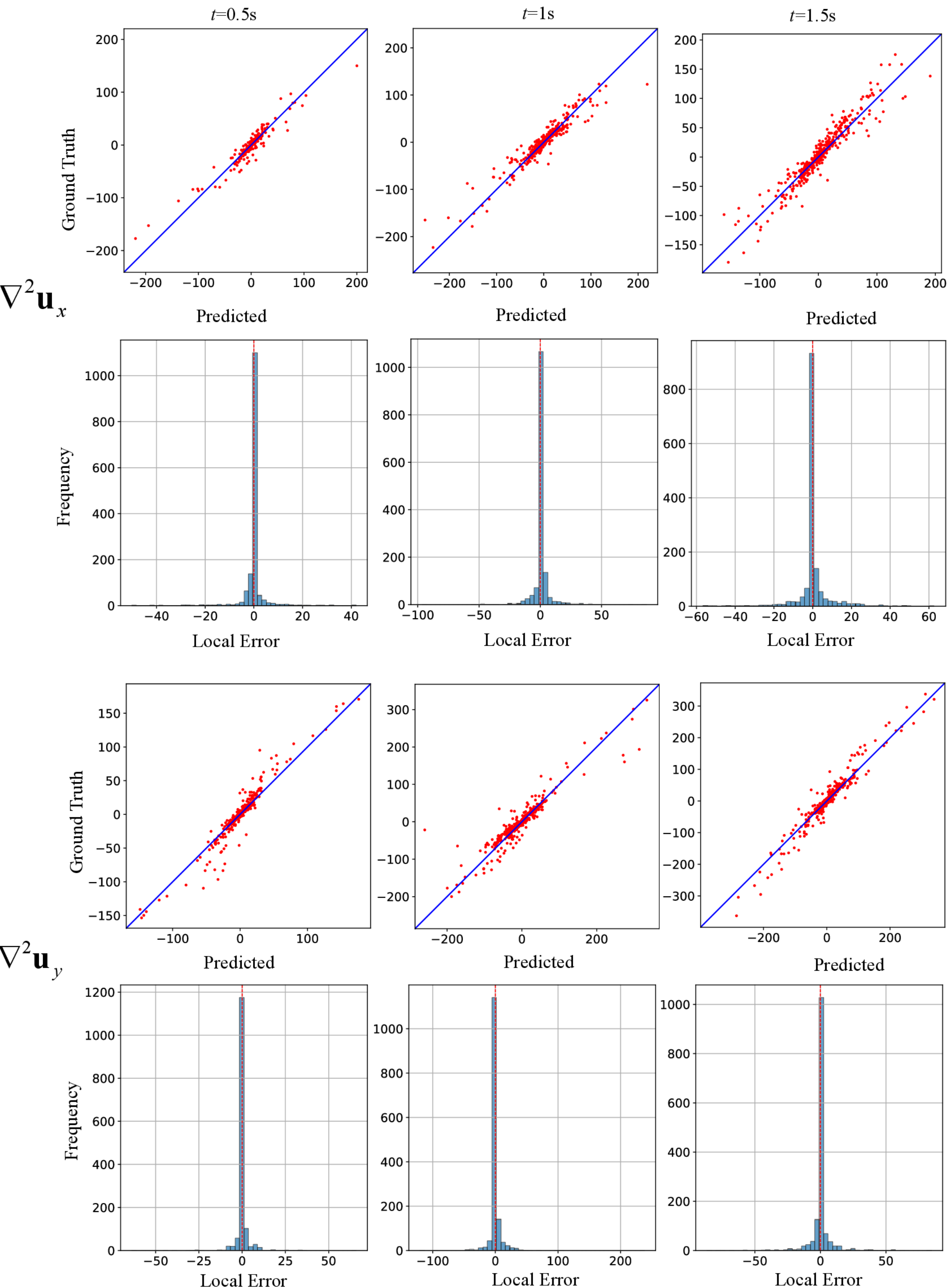}
  \caption[width=0.7\textwidth]{Parity plots (top row) and error histograms (bottom row) comparing predicted boundary contributions $\hat{B}_i$ with ground truth values ${B}_i$ for laplacian of velocity ($\nabla^2 \boldsymbol{u}_x$, $\nabla^2 \boldsymbol{u}_y$) in Case 3}
  \label{Fig:adv-diff laplacianVelocity for Histogram and errors}
\end{figure}

\begin{figure}[H]
  \centering
  \includegraphics[width=0.9\textwidth]{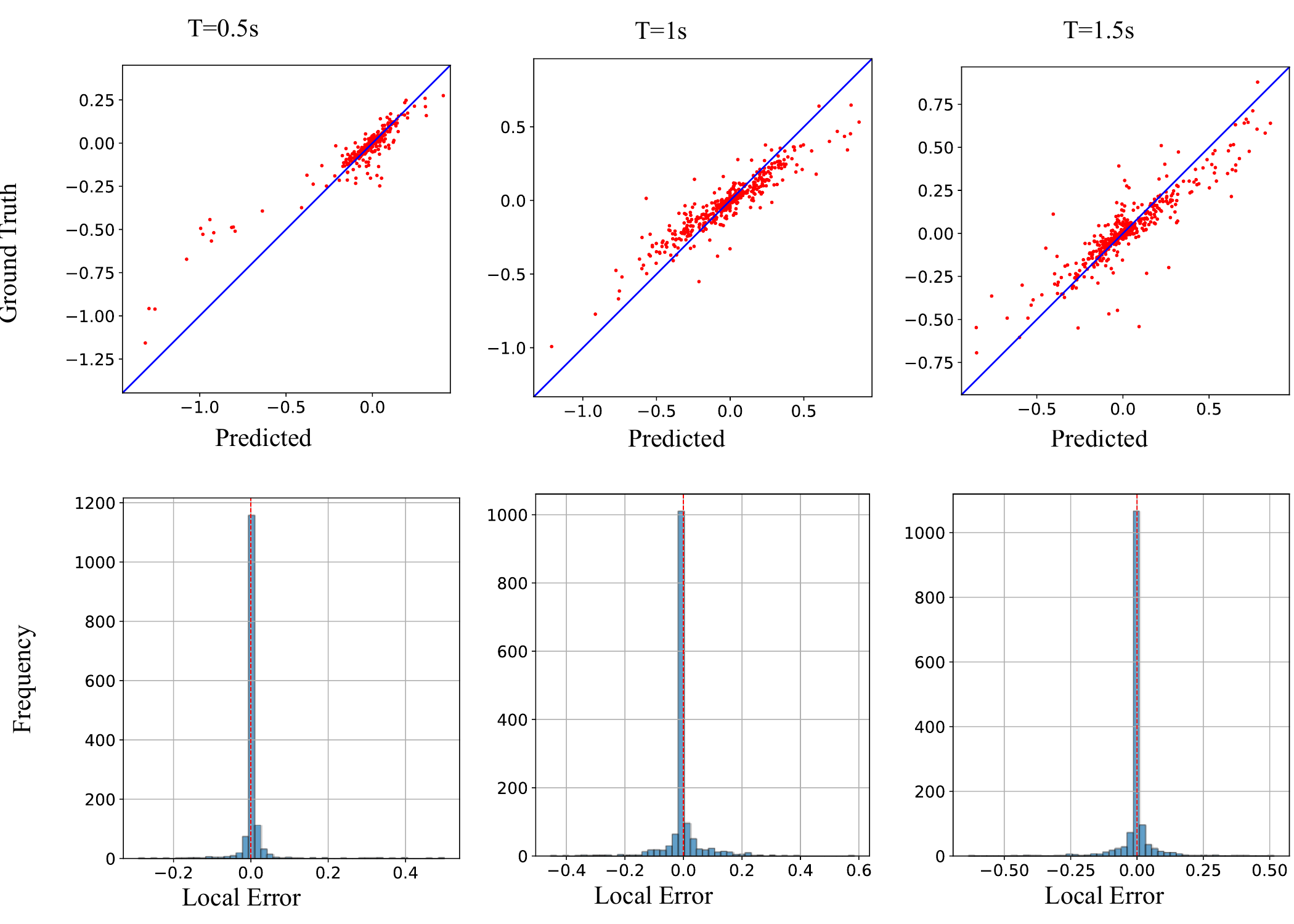}
  \caption[width=0.7\textwidth]{Parity plots (top row) and error histograms (bottom row) comparing predicted boundary contributions $\hat{B}_i$ with ground truth values ${B}_i$ for  velocity divergence ($\nabla \cdot \boldsymbol{u}$)  in Case 3}
  \label{Fig:adv-diff divergenceVelocity for Histogram and errors}
\end{figure}

\begin{figure}[H]
  \centering
  \includegraphics[width=.9\textwidth]{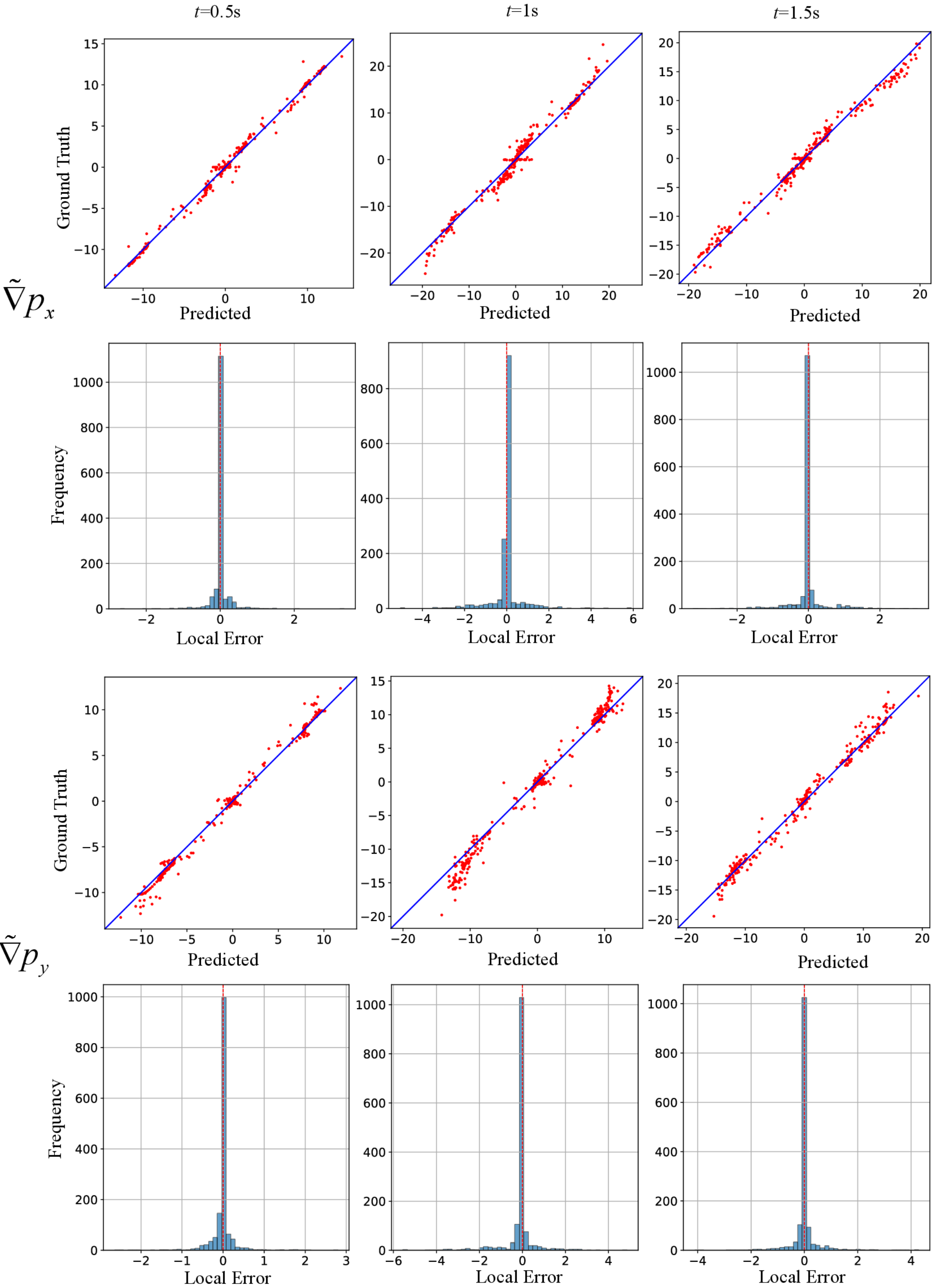}
  \caption[width=0.7\textwidth]{Parity plots (top row) and error histograms (bottom row) comparing predicted boundary contributions $\hat{B}_i$ with ground truth values ${B}_i$ for pressure gradient ($\nabla p_x$, $\nabla p_y$)  in Case 1}
  \label{Fig:adv-diff gradientPressure for Histogram and errors}
\end{figure}

The parity plots in Figures \ref{Fig:adv-diff_n0_for_Histogram_and_errors}, \ref{Fig:adv-diff laplacianTemprature for Histogram and errors}, \ref{Fig:adv-diff laplacianVelocity for Histogram and errors}, \ref{Fig:adv-diff divergenceVelocity for Histogram and errors}, \ref{Fig:adv-diff gradientPressure for Histogram and errors} support these observations. Predictions for first-order quantities, such as the pressure gradient, align almost perfectly with the ground truth, showing minimal scatter. In contrast, second-order quantities—especially the velocity Laplacians, display a wider spread and systematic deviations at larger magnitudes, consistent with the higher NRMSE values reported in Table\ref{table:adv-diffErrors}. Nonetheless, the strong clustering of points along the diagonal demonstrates that the model generalizes effectively and maintains accuracy even for complex derivative quantities.

\begin{table}[h]
\centering
\caption{Performance metrics: only $n$ is available for $t=0.0\,\mathrm{s}$; other variables are for $t=0.5\,\mathrm{s}$, and $t=1.0\,\mathrm{s}$.}
\resizebox{\textwidth}{!}{
\begin{tabular}{l
r
rrrrrrr
rrrrrrr}
\toprule
 & \multicolumn{1}{c}{$t=0.0\,\mathrm{s}$}
 & \multicolumn{7}{c}{$t=0.5\,\mathrm{s}$}
 & \multicolumn{7}{c}{$t=1.0\,\mathrm{s}$} \\
\cmidrule(lr){2-2} \cmidrule(lr){3-9} \cmidrule(lr){10-16}
 & $n$
 & $n$ & $\nabla^2 T$ & $\tilde\nabla p_x$ & $\tilde\nabla p_y$ & $\nabla\cdot\boldsymbol{u}$ & $\nabla^2\boldsymbol{u}_x$ & $\nabla^2\boldsymbol{u}_y$
 & $n$ & $\nabla^2 T$ & $\tilde\nabla p_x$ & $\tilde\nabla p_y$ & $\nabla\cdot\boldsymbol{u}$ & $\nabla^2\boldsymbol{u}_x$ & $\nabla^2\boldsymbol{u}_y$ \\
\midrule
R        
& 1.000
& 0.993 & 0.987 & 0.997 & 0.997 & 0.942 & 0.952 & 0.956
& 0.993 & 0.987 & 0.989 & 0.990 & 0.927 & 0.950 & 0.941 \\
NMSE(\%)  
& 0.192
& 2.299 & 0.584 & 0.358 & 0.407 & 0.779 & 0.431 & 0.469
& 1.180 & 0.474 & 0.525 & 0.673 & 1.516 & 0.787 & 0.733 \\
NRMSE (\%) 
& 0.556
& 4.167 & 1.910 & 0.974 & 1.088 & 2.923 & 1.376 & 1.568
& 2.651 & 1.581 & 1.398 & 1.774 & 3.763 & 2.229 & 2.242 \\
\bottomrule
\end{tabular}}

\label{table:adv-diffErrors}
\end{table}

\section{Conclusion}

We presented a physics-informed machine learning (ML) framework for computing solid boundary contributions in particle methods. The approach learns from the ghost-particle method to predict correction terms in each MPS approximation operator, thereby eliminating the need for explicit ghost particles. The architecture combines convolutional and fully connected layers to process physics-inspired features, including geometry, field variables, and kernel properties, into accurate boundary predictions. Training on datasets with diverse geometries and field conditions enabled the model to generalize across a wide range of scenarios.

The results demonstrate that the hybrid CNN–MLP models are accurate, robust, and broadly applicable for boundary treatment in particle discretizations. They consistently achieved near-perfect correlations with ghost-particle results, while avoiding overfitting and maintaining high accuracy ($R > 0.94$, $NRMSE < 4\%$) even for challenging second-order derivatives. Importantly, the models generalized well to unseen geometries, varying spatial frequencies, non-uniform particle distributions, and unsteady Navier–Stokes flows. Errors were largely confined to geometric singularities, and performance remained stable across both static and dynamic conditions, suggesting strong potential for use in other particle-based solvers without retraining.

This study focused on two-dimensional problems, where the computational cost of ML and ghost-particle methods is comparable. Future work will extend the framework to three dimensions, where eliminating ghost particles can offer substantial efficiency gains. We also plan to explore adaptive boundary representations using nodes or polygonal meshes (instead of wall particles), enabling local resolution control and further reductions in computational cost. Beyond MPS, the framework can be directly applied to SPH and related meshfree methods, where boundary treatment often dominates accuracy. Finally, incorporating a richer set of dimensionless, scale-independent features may enhance generalization to flows at very different scales.

\section{CRediT authorship contribution statement}
\textbf{Nariman Mehranfar}: Writing – original draft, Validation, Software, Methodology, Formal analysis, Data curation, Conceptualization.
\textbf{Ahmad Shakibaeinia}: Writing – review \& editing, Supervision, Project administration, Methodology, Investigation, Funding acquisition, Formal analysis, Data curation, Conceptualization. 

\section{Declaration of competing interest}
The authors declare that they have no known competing financial interests or personal relationships that could have appeared to influence the work reported in this paper.

\section{Declaration of generative AI and AI-assisted technologies in the writing process}
During the preparation of this work, the authors used Grammarly and ChatGPT to check grammatical correctness and to improve the readability and clarity of the text. After using these tools, the authors reviewed and edited the content as necessary and take full responsibility for the final content of the published article.

\section{Acknowledgments}
This work was funded by the Canada Research Chair (CRC) Program. Computational infrastructure was provided through the John R. Evans Leaders Fund (JELF) of the Canada Foundation for Innovation (CFI).

\appendix
\renewcommand{\thefigure}{(\Alph{section}-\arabic{figure})}

\section{Example simulation results of Case 3}
\label{app1}
\setcounter{figure}{0} 

\begin{figure}[H]
  \centering
  \includegraphics[width=0.75\textwidth]{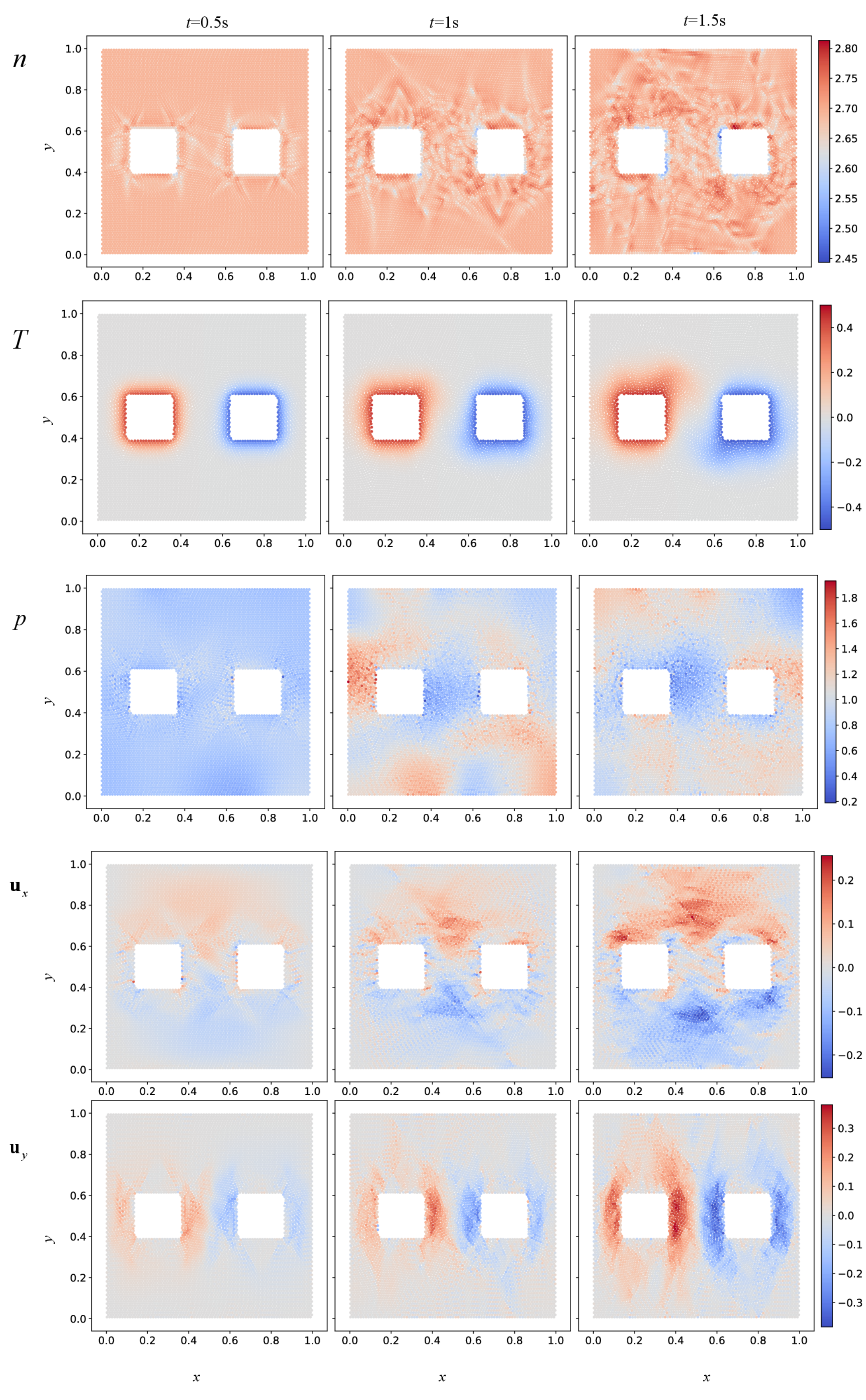}
  \caption{MPS-simulated particle number density ($n$), temperature ($T$), pressure ($p$), and velocity field $\mathbf{u}=(u_x, u_y)$ at different time steps for Case~3.}
  \label{Fig:case3all}
\end{figure}

Figure~\ref{Fig:case3all} presents snapshots of the simulated flow-field variables for Case 3, as obtained from the MPS solution.

\section{Cross-validation trade-off for training NN}
\label{app2}
\setcounter{figure}{0} 

\begin{figure}[H]
  \centering
  \includegraphics[width=\textwidth]{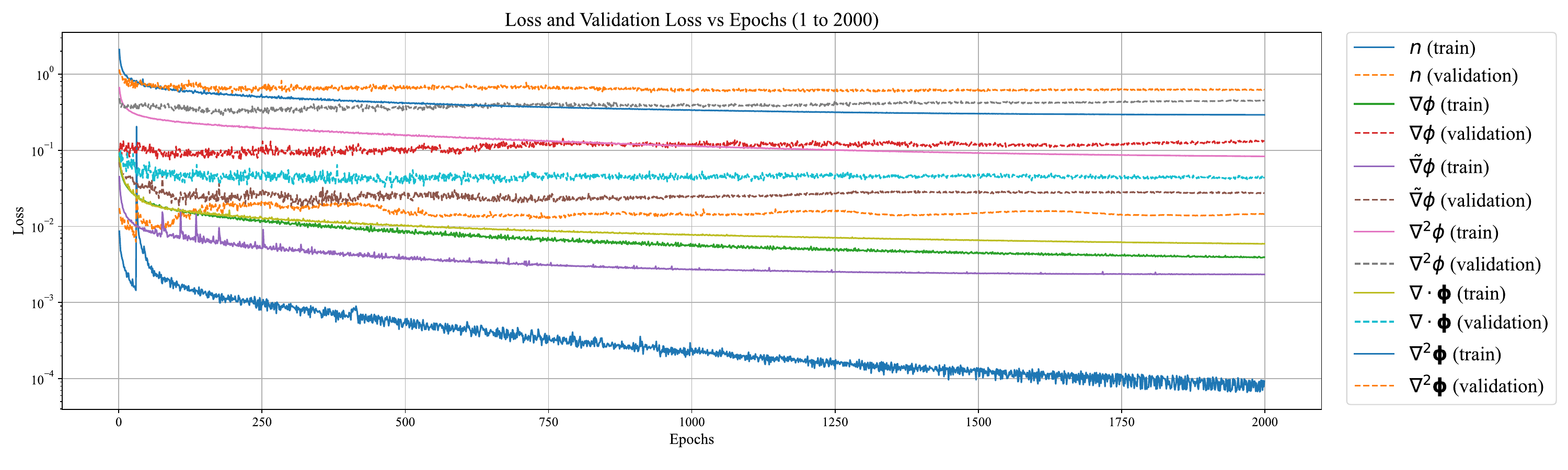}
  \caption{Evolution of the loss function for training and validation datasets
  over 2000 epochs.}
  \label{Fig:loss}
\end{figure}

The validation and training loss values are illustrated in Fig.~\ref{Fig:loss}.

\bibliographystyle{elsarticle-num-names}
\bibliography{references.bib}

\end{document}